\documentclass[twocolumn,times]{aastex62}
\usepackage{graphicx}
\usepackage[flushleft]{threeparttable}
\usepackage{blindtext}
\usepackage{amsmath}
\usepackage{mathtools}
\usepackage{multirow}
\usepackage{comment}
\turnoffeditone

\DeclareRobustCommand{\ion}[2]{%
\relax\ifmmode
\ifx\testbx\f@series
{\mathbf{#1\,\mathsc{#2}}}\else
{\mathrm{#1\,\mathsc{#2}}}\fi
\else\textup{#1\,{\mdseries\textsc{#2}}}%
\fi}

\begin{document}
\shortauthors{Jung et al.}
\def\nar{New Astron.}
\def\na{New Astron.}
\title{\large \textbf{New \textit{z} $>$ 7 Lyman-alpha Emitters in EGS: Evidence of an Extended Ionized Structure at \textit{z} $\sim$ 7.7}}

\author[0000-0003-1187-4240]{Intae Jung}
\affil{Space Telescope Science Institute, 3700 San Martin Drive Baltimore, MD 21218, USA}
\affil{Astrophysics Science Division, NASA Goddard Space Flight Center, 8800 Greenbelt Rd, Greenbelt, MD 20771, USA}
\affil{Department of Physics, The Catholic University of America, Washington, DC 20064, USA }

\author[0000-0001-8519-1130]{Steven L. Finkelstein}
\affiliation{Department of Astronomy, The University of Texas at Austin, Austin, TX, USA}

\author[0000-0003-2366-8858]{Rebecca L. Larson}
\altaffiliation{NSF Graduate Fellow}
\affiliation{Department of Astronomy, The University of Texas at Austin, Austin, TX, USA}

\author[0000-0001-6251-4988]{Taylor A. Hutchison}
\altaffiliation{NASA Postdoctoral Fellow}
\affiliation{Astrophysics Science Division, NASA Goddard Space Flight Center, 8800 Greenbelt Rd, Greenbelt, MD 20771, USA}

\author[0000-0002-4772-7878]{Amber N. Straughn}
\affiliation{Astrophysics Science Division, NASA Goddard Space Flight Center, 8800 Greenbelt Rd, Greenbelt, MD 20771, USA}

\author[0000-0002-9921-9218]{Micaela B. Bagley}
\affiliation{Department of Astronomy, The University of Texas at Austin, Austin, TX, USA}

\author[0000-0001-9875-8263]{Marco Castellano}
\affiliation{INAF - Osservatorio Astronomico di Roma, via di Frascati 33, 00078 Monte Porzio Catone, Italy}

\author[0000-0001-7151-009X]{Nikko J. Cleri}
\affiliation{Department of Physics and Astronomy, Texas A\&M University, College Station, TX, 77843-4242 USA}
\affiliation{George P.\ and Cynthia Woods Mitchell Institute for Fundamental Physics and Astronomy, Texas A\&M University, College Station, TX, 77843-4242 USA}

\author[0000-0003-1371-6019]{M. C. Cooper}
\affiliation{Department of Physics \& Astronomy, University of California, Irvine, 4129 Reines Hall, Irvine, CA 92697, USA}

\author[0000-0001-5414-5131]{Mark Dickinson}
\affiliation{NSF's National Optical-Infrared Astronomy Research Laboratory, 950 N. Cherry Ave., Tucson, AZ 85719, USA}

\author[0000-0001-7113-2738]{Henry C. Ferguson}
\affil{Space Telescope Science Institute, 3700 San Martin Drive Baltimore, MD 21218, USA}

\author[0000-0002-4884-6756]{Benne W. Holwerda}
\affil{Physics \& Astronomy Department, University of Louisville, 40292 KY, Louisville, USA}

\author[0000-0001-9187-3605]{Jeyhan S. Kartaltepe}
\affiliation{Laboratory for Multiwavelength Astrophysics, School of Physics and Astronomy, Rochester Institute of Technology, 84 Lomb Memorial Drive, Rochester, NY 14623, USA}

\author[0000-0002-6787-3020]{Seonwoo Kim}
\affiliation{Department of Astronomy, University of Illinois, 1002 West Green Street, Urbana, IL 61801, USA}

\author[0000-0002-6610-2048]{Anton M. Koekemoer}
\affil{Space Telescope Science Institute, 3700 San Martin Drive Baltimore, MD 21218, USA}

\author[0000-0001-7503-8482]{Casey Papovich}
\affiliation{Department of Physics and Astronomy, Texas A\&M University, College Station, TX, 77843-4242 USA}
\affiliation{George P.\ and Cynthia Woods Mitchell Institute for Fundamental Physics and Astronomy, Texas A\&M University, College Station, TX, 77843-4242 USA}

\author[0000-0002-7464-7857]{Hyunbae Park}
\affiliation{Lawrence Berkeley National Laboratory, CA 94720, USA}
\affiliation{Berkeley Center for Cosmological Physics, UC Berkeley, CA 94720, USA}

\author[0000-0001-8940-6768]{Laura Pentericci}
\affiliation{INAF - Osservatorio Astronomico di Roma, via di Frascati 33, 00078 Monte Porzio Catone, Italy}

\author[0000-0003-4528-5639]{Pablo G. P\'erez-Gonz\'alez}
\affiliation{Centro de Astrobiolog\'{\i}a (CAB), CSIC-INTA, Ctra. de Ajalvir km 4, Torrej\'on de Ardoz, E-28850, Madrid, Spain}

\author[0000-0002-8442-3128]{Mimi Song}
\affiliation{Department of Astronomy, University of Massachusetts, Amherst, MA, 01002, USA}

\author[0000-0002-8224-4505]{Sandro Tacchella}
\affiliation{Kavli Institute for Cosmology, University of Cambridge, Madingley Road, Cambridge, CB3 0HA, UK}
\affiliation{Cavendish Laboratory, University of Cambridge, 19 JJ Thomson Avenue, Cambridge, CB3 0HE, UK}

\author[0000-0001-6065-7483]{Benjamin J. Weiner}
\affiliation{MMT/Steward Observatory, University of Arizona, 933 N. Cherry Ave., Tucson, AZ 85721, USA}

\author[0000-0001-9262-9997]{Christopher N. A. Willmer}
\affiliation{Steward Observatory, University of Arizona, 933 N. Cherry Ave., Tucson, AZ, 85721, USA}

\author[0000-0002-7051-1100]{Jorge A. Zavala}
\affiliation{National Astronomical Observatory of Japan, 2-21-1 Osawa, Mitaka, Tokyo 181-8588, Japan}

\correspondingauthor{Intae Jung}
\email{ijung@stsci.edu}

\submitjournal{the Astrophysical Journal}

\begin{abstract}
We perform a ground-based near-infrared spectroscopic survey using the Keck/MOSFIRE spectrograph to target Ly$\alpha$ emission at $7.0<z<8.2$ from 61 galaxies to trace the ionization state of the intergalactic medium (IGM). We cover a total effective sky area of $\sim10^\prime\times10^\prime$ in the Extended Groth Strip field of the Cosmic Assembly Near-infrared Deep Extragalactic Legacy Survey. From our observations, we detect Ly$\alpha$ emission at a $>$4$\sigma$ level in eight $z>7$ galaxies, which include additional members of the known $z\sim7.7$ Ly$\alpha$-emitter (LAE) cluster \citep{Tilvi2020a}. With the addition of these newly-discovered $z\sim7.7$ LAEs, this is currently the largest measured LAE cluster at $z>7$. The unusually-high Ly$\alpha$ detection rate at $z\sim7.7$ in this field suggests significantly stronger Ly$\alpha$ emission from the clustered LAEs than from the rest of our targets. We estimate the ionized bubble sizes around these LAEs and conclude that the LAEs are clustered within an extended ionized structure created by overlapping ionized bubbles which allow the easier escape of Ly$\alpha$ from galaxies. It is remarkable that the brightest object in the cluster has the lowest measured redshift of the Ly$\alpha$ line, being placed in front of the other LAEs in the line-of-sight direction. This suggests that we are witnessing the enhanced IGM transmission of Ly$\alpha$ from galaxies on the rear side of an ionized area. This could be a consequence of Ly$\alpha$ radiative transfer: Ly$\alpha$ close to the central velocity is substantially scattered by the IGM while Ly$\alpha$ from the rear-side galaxies is significantly redshifted to where it has a clear path.
\end{abstract}

\section{Introduction}
Investigating the ionization state of the intergalactic medium (IGM) during the epoch of reionization is critical to understanding the formation and evolution of galaxies in the early Universe. Along with contributions from active galactic nucleus (AGN) activity \citep[e.g.,][]{Matsuoka2018a, Kulkarni2019a, Dayal2020a}, galaxies are responsible for supplying the bulk of ionizing photons into the IGM at early cosmic time \citep[e.g.,][]{Robertson2015a, McQuinn2016a, Dayal2018a, Finkelstein2019a, Robertson2021a}.

Lyman-alpha (Ly$\alpha$) emission has been used as an observational probe of the ionization state of the IGM during the epoch of reionization \citep[e.g.,][]{Miralda-Escude1998a, Rhoads2001a, Stark2011a, Pentericci2011a, Dijkstra2014a}.  A rapid decline in the Ly$\alpha$ fraction\footnote{Ly$\alpha$ fraction is defined as $N_\text{LAE}/N_\text{LBG}$, where $N_\text{LAE}$ is the number of Ly$\alpha$-detected objects and $N_\text{LBG}$ is the number of high-redshift-candidate Lyman-Break Galaxies (LBGs) observed in spectroscopic observations.} at $z>6$ suggests that the Ly$\alpha$ visibility is strongly affected by the IGM attenuation into the epoch of reionization \citep[extensively reviewed by][and the references therein]{Ouchi2020a} while the evolutionary effect of host galaxy properties could impact the observed evolution of Ly$\alpha$ \citep[e.g.,][]{Mesinger2015a, Hassan2021a}.

Thanks to the infrared (IR) wavelength coverage of JWST, it has become possible to deliver spectroscopic confirmations of reionization-era galaxies by detecting additional emission lines, which -- in contrast to Ly$\alpha$ -- are not affected by the neutral IGM \citep[e.g.,][]{Brinchmann2022a, Schaerer2022a, Trump2022a, Trussler2022a}.  As expected, Ly$\alpha$ emission has not been detected from the recent JWST NIRSpec observations of $z>9$ galaxies \citep{Roberts-Borsani2022b, Williams2022a, Curtis-Lake2022a, Wang2022a}.  This suggests that the sizes of ionized bubbles around these galaxies might not yet be sufficiently large enough to allow for the escape of Ly$\alpha$, and their rapid growth has not yet occurred in this early stage of reionization.  

At later stages of reionization, Ly$\alpha$ may become increasingly visible as ionized bubbles around galaxies grow over time. While a dearth of Ly$\alpha$ emission detected at $z>8$ implies a significantly-neutral IGM in the early stage of reionization \citep[a handful of detections reported in][]{Zitrin2015a, Laporte2017a, Larson2022a}, a significant number of Ly$\alpha$-emission lines have been detected at $z\gtrsim7$, preferentially in UV-luminous galaxies \citep{Oesch2015a, Roberts-Borsani2016a, Zheng2017a, Castellano2018a, Tilvi2020a, Jung2020a, Hu2021a, Jung2022a, Endsley2021b, Endsley2022a}.  Thus, Ly$\alpha$ observations in the middle/late phases of reionization play a key role in tracing the evolution of ionized structures in the IGM.

Specifically, spectroscopic searches for Ly$\alpha$ in the middle phase of reionization at $z\sim$ 7 -- 8 provide a higher detection rate of Ly$\alpha$ particularly from UV-brighter galaxies \citep[e.g.,][and references mentioned above]{Jung2022a}, compared to rarer detections from fainter ones \citep{Hoag2019a, Roberts-Borsani2022a}.  This may indicate an inhomogeneous process of reionization where ionizing photons from UV-luminous galaxies in overdense regions are likely to ionize the IGM around them earlier than isolated UV-fainter galaxies \citep[e.g.,][]{Mesinger2011a, Ocvirk2021a, Kannan2022a}.  A continuing effort for Ly$\alpha$ observations is necessary to capture the global evolution of reionization, probing volumes larger than local ionized structures. 

In this paper, we present new spectroscopic observations of reionization-era galaxies. Our study provides spectral coverage for Ly$\alpha$ emission from a large number of high-redshift candidate galaxies in a section of the Cosmic Assembly Near-infrared Deep Extragalactic Legacy Survey \citep[CANDELS][]{Grogin2011a, Koekemoer2011a} Extended Groth Strip (EGS) field, with a total effective area of $\sim10^\prime\times10^\prime$.  Our spectroscopic observations deliver new Ly$\alpha$ emission lines detected from $z>7$ galaxies, uncovering the largest LAE cluster\footnote{To clarify, our discussion on LAE clusters must be distinguished from the conventional definition of galaxy clusters in the context of forming virialized systems. Instead, we discuss LAE clusters whose LAEs overlap individual ionized bubbles each other, forming contiguous ionized areas.} system in this early Universe at $z>7$.  The observations suggest that there is an extended ionized structure associated with the clustered LAEs, which enhances the transmission of Ly$\alpha$ along our line of sight. Non-detections of Ly$\alpha$ from the bulk of our targets reinforce earlier indications that the IGM at $z>7$ is on average more neutral than at lower redshifts.

This paper is structured as follows. In Section 2, we describe our spectroscopic targets, MOSFIRE observations, and data reduction.  We present the Ly$\alpha$-emission lines detected in our observations, giving the measured physical properties of these emission lines and their host galaxies in Section 3. Section 4 discusses the extended ionized structure around the clustered LAEs at $z\sim7.7$ in the EGS field. We then summarize our findings in Section 5.  In this work, we assume the Planck cosmology \citep{Planck-Collaboration2016a} with $H_0$ = 67.8\,km\,s$^{-1}$\,Mpc$^{-1}$, $\Omega_{\text{M}}$ = 0.308, and $\Omega_{\Lambda}$ = 0.692. We use pMpc to indicate proper distances and cMpc to indicate co-moving distances. The Hubble Space Telescope (HST) F606W, F814W, F105W, F125W, F140W, and F160W bands are referred to as $V_{606}$, $I_{814}$, $Y_{105}$, $J_{125}$, $JH_{140}$ and $H_{160}$, respectively.  All magnitudes in this work are quoted in the AB system \citep{Oke1983a}, and all errors mentioned in this paper represent 1$\sigma$ uncertainties (or central 68\% confidence ranges) unless stated otherwise.

\section{Data}
\subsection{Targets}
Targets were selected from the photometrically-selected high-redshift galaxy catalog of \cite{Finkelstein2022a}, which is based on the updated HST CANDELS photometry. The photometric selection of high-redshift galaxies is done as described in Section 3.2 in \cite{Finkelstein2015a}, using the photometric redshift ($z_p$) probability distribution functions (PDFs) of $P(z)$ calculated by {\sc EAZY} \citep{Brammer2008a}.
Then, we created a target list of $J_{\text{125}}\lesssim27$ galaxies with $z_p>6$ in the CANDELS/EGS field, which was used for designing optimized slitmask configurations in MAGMA\footnote{https://www2.keck.hawaii.edu/inst/mosfire/magma.html} for our Keck/MOSFIRE observations.  In our slitmask design, we prioritized targets on slits based on the galaxy brightness ($J_{\text{125}}$) and the integrals of $P(z)$ in $7.0<z<8.2$, which corresponds to the MOSFIRE $Y$-band wavelength coverage for Ly$\alpha$ emission. This resulted in 61 Ly$\alpha$ targets across our four MOSFIRE pointings. 

\subsection{Photometric Data and Galaxy Properties}
We use the photometric catalog of \cite{Finkelstein2022a}, which includes the HST ACS and WFC3 broadband photometry ($V_{606}$, $I_{814}$, $Y_{105}$, $J_{125}$, $JH_{140}$, and $H_{160}$) in addition to Spitzer/IRAC 3.6$\mu$m and 4.5$\mu$m band fluxes in the CANDELS/EGS field.  We also use photometric redshift measurements that have been obtained with {\sc EAZY} in \cite{Finkelstein2022a} based on the updated CANDELS photometry. 

To derive galaxy physical properties, we performed spectral energy distribution (SED) fitting with the photometric data to galaxy SED models.  In the construction of galaxy model SEDs, we assume a \cite{Salpeter1955a} initial mass function with a stellar mass range of 0.1-100$M_{\odot}$.  We allow a range of metallicity from 0.005$Z_{\odot}$ to 1.0$Z_{\odot}$, and exponential models of star formation histories are used with exponentially varying timescales, parameterized with $\tau=$ 10 Myr, 100 Myr, 1 Gyr, 10 Gyr, 100 Gyr, $-$300 Myr, $-$1 Gyr, $-$10 Gyr.  We use the \cite{Calzetti2001a} dust attenuation description for a ranging from 0 to 0.8 mag in $E(B-V)$ values.  Nebular emission lines are added, based on the \cite{Inoue2011a} emission-line ratio, through the same process as done in \cite{Salmon2015a}.  The IGM attenuation was applied to model the galaxy SEDs, following \cite{Madau1995a}. 

Fiducial values of SED-derived physical properties, such as stellar masses, the absolute UV magnitudes ($M_{\text{UV}}$), and the UV continuum slope ($\beta$), were obtained from the best-fit models, which minimize $\chi^2$ to the observed photometry.  We estimated the uncertainties of physical quantities from SED fitting with 1000 Monte Carlo (MC) realizations of the simulated photometric fluxes, which we perturbed the observed fluxes with their photometric errors. The 1$\sigma$ uncertainties denote the upper and lower limits of the central 68\% range taken from the 1000 MC simulations. We repeated the process for all individual targets.  We fixed galaxy redshifts with Ly$\alpha$-derived spectroscopic redshifts for emission-detected objects and with the best-fit photometric redshifts for non-detection objects.  We derived $M_{\text{UV}}$ by averaging fluxes over a 100\AA-bandpass (at the rest-frame 1450 -- 1550\AA) from SED models, which are not dust-corrected.  The rest-frame UV continuum ($\beta$) was measured in the rest-frame UV bandpass of 1300 -- 2600\AA\ from the best-fit SED models as well, where $\beta$ is the spectral index in the form of $f_{\lambda}\propto\lambda^{\beta}$. 

\begin{figure}[t]
\centering
\includegraphics[width=1.0\columnwidth]{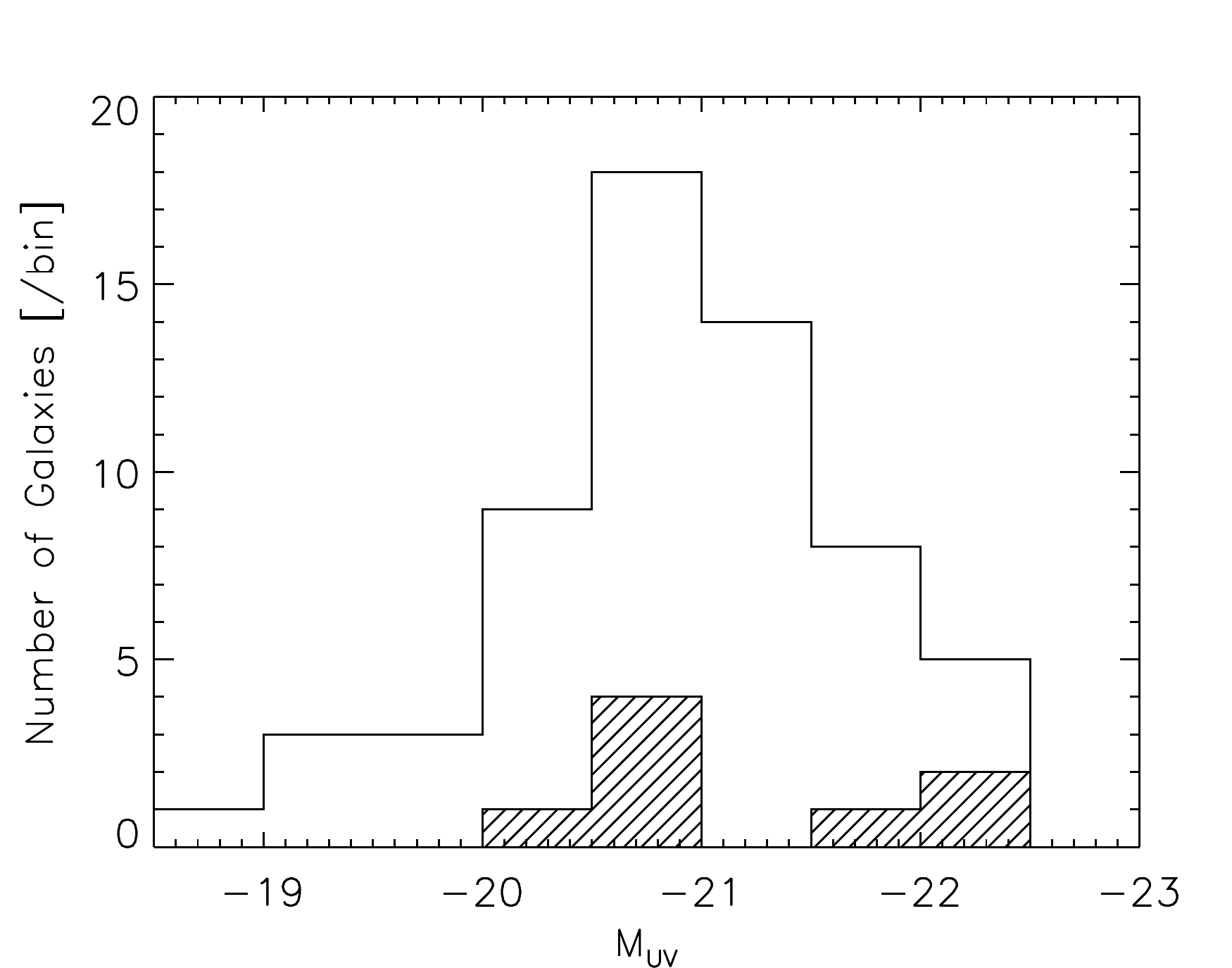}
\caption{$M_{\text{UV}}$ distribution of galaxies targeted in our MOSFIRE observations. A majority of the targets ($>$90\%) are relatively UV bright with $M_{\text{UV}}\lesssim-20$, which is comparable to the detection limit of the CANDELS/HST $J_{125}$ imaging depth in the EGS field.  The shaded histogram indicates the $M_{\text{UV}}$ of the sources with new Ly$\alpha$ detections.}
\label{fig:muv}
\end{figure}

\begin{figure}[t]
\centering
\includegraphics[width=1.0\columnwidth]{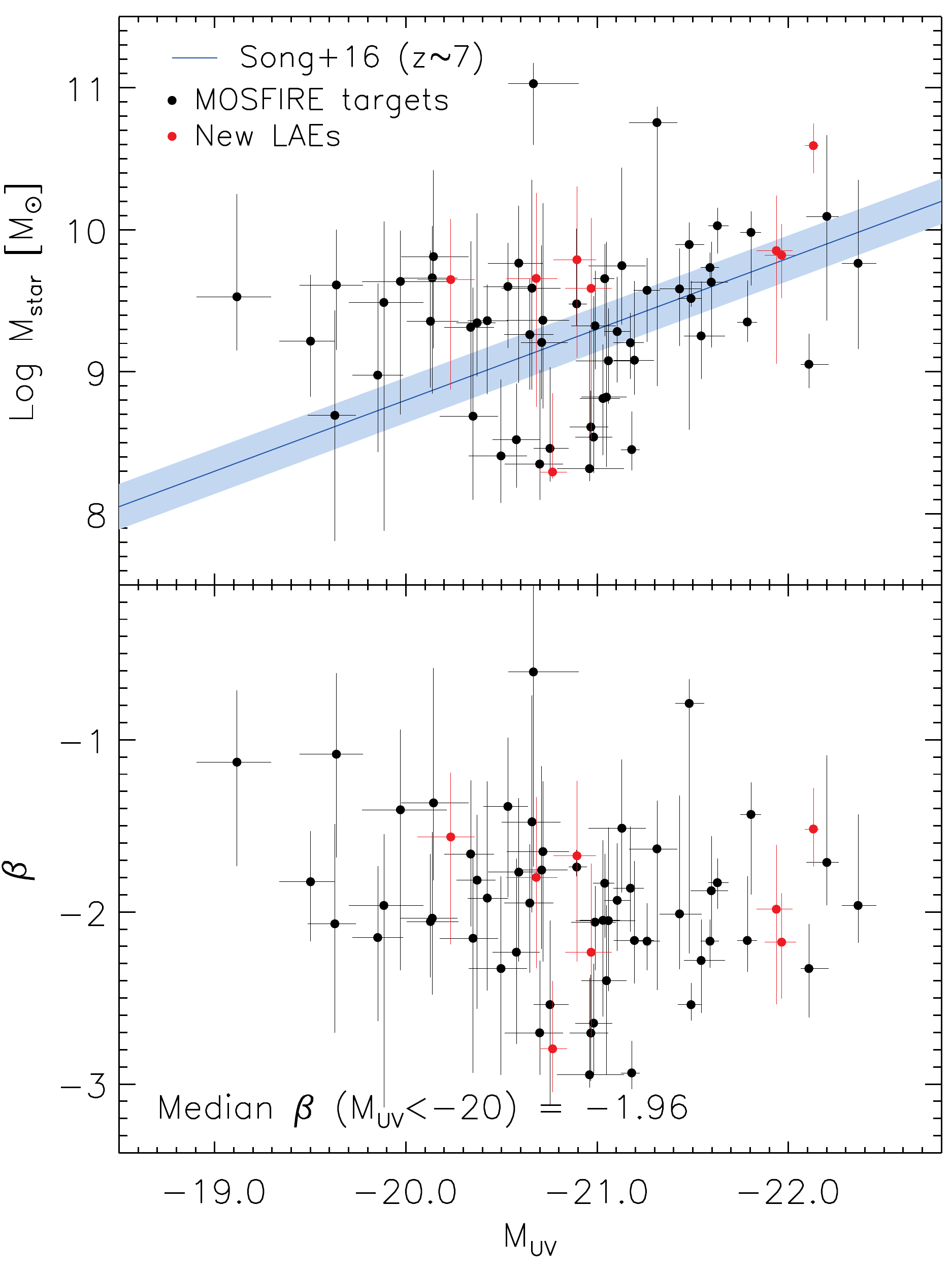}
\caption{(Top) Our spectroscopic targets presented in the $M_\text{star}$--$M_\text{UV}$ plane. The blue line with shaded region shows the $z\sim7$ $M_\text{star}$--$M_\text{UV}$ relation in \cite{Song2016a}. (Bottom) The measurements of the rest-UV continuum slope ($\beta$) versus $M_\text{UV}$. The red symbols represent Ly$\alpha$-detected targets.}
\label{fig:muv_m_beta}
\end{figure}

We present the $M_\text{UV}$ distribution of our targets in Figure \ref{fig:muv}. Our targets are somewhat biased toward UV-brighter galaxies, and a majority of the targets have $M_{\text{UV}}\lesssim-20$, comparable to the CANDELS/HST $J_{125}$ imaging depth in the EGS field.  Figure \ref{fig:muv_m_beta} presents our targets in the $M_\text{star}$--$M_\text{UV}$ plane (top) and their rest-UV continuum slope ($\beta$) versus $M_\text{UV}$ (bottom).  Although our sample contains limited coverage of UV-faint ($M_\text{UV}>-20$) sources, our spectroscopic targets are broadly consistent with the  $z\sim7$ $M_\text{star}$--$M_\text{UV}$ relation \citep{Song2016a}, which is representative of the typical high-redshift galaxy population.  Also, we find the median value of the rest-UV continuum slopes at $\beta=-1.96$ from $M_\text{UV}<-20$ galaxies.  This is comparable to the typical range of the UV slope measurements at this redshift (e.g., \citealt{Finkelstein2012a} find $\beta=-2.15^{+0.25}_{-0.16}$ for $M_\text{UV}<-20$ at $z=7$).

\begin{figure*}[ht]
\centering
\includegraphics[width=0.83\paperwidth]{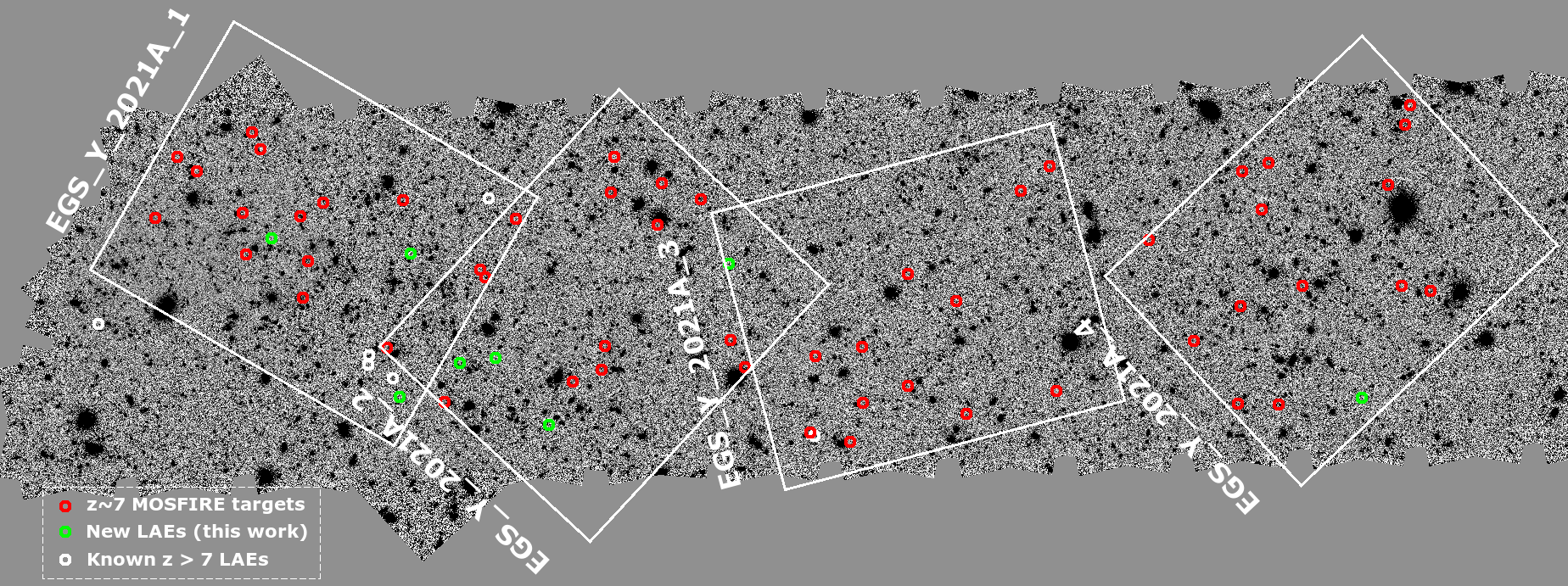}
\caption{Four MOSFIRE mask configurations (white rectangles) overlaid in the CANDELS EGS $H_{160}$-band image. Our high-redshift spectroscopic targets for Ly$\alpha$ are shown as red circles, and the newly-detected $z>7$ LAEs in this work are marked as green circles. We show the known $z>7$ LAEs in EGS as white circles \citep{Zitrin2015a, Oesch2015a, Roberts-Borsani2016a, Tilvi2020a, Larson2022a}.}
\label{fig:masks}
\end{figure*}

\begin{deluxetable*}{cccccccc}
\tablecaption{Summary of Keck/MOSFIRE Observations in EGS\label{tab:observations}} 
\tablehead{
\colhead{Mask Name} & \colhead{R.A. (J2000.0)} &\colhead{Decl. (J2000.0)} & \colhead{Observational Date} & \colhead{$N_\text{targets}$}  & \colhead{$t_\text{exp}$} & \colhead{Seeing\tablenotemark{a}} & \colhead{Standard Star\tablenotemark{b}} \\
	\colhead{}  & \colhead{(degree)}  & \colhead{(degree)}  & \colhead{}  & \colhead{} & \colhead{(hr)} & \colhead{(arcsec)} & \colhead{} }
\startdata
{EGS\_Y\_2021A\_1} & {215.11787} & {53.03937} & {2021 Apr 23} & {17} & {3.5} & {0.7} & {HIP56147}\\
{EGS\_Y\_2021A\_2} & {215.05683} & {52.95982} & {2021 Apr 23} & {16} & {3.2} & {0.9} & {HIP56147}\\
{EGS\_Y\_2021A\_3} & {214.95563} & {52.89208} & {2021 Apr 24} & {13} & {3.6} & {1.2} & {HIP56147}\\
{EGS\_Y\_2021A\_4} & {214.80996} & {52.80919} & {2021 Apr 24} & {15} & {3.4} & {1.0} & {HIP56147}\\
\enddata
\tablenotetext{}{
$^{a}$\footnotesize Full-width half maximum (FWHM) estimated from continuum objects in science mask configurations.\\
$^{b}$Standard star in our long-slit observations for flux calibration, listed in the \textit{Hipparcos} index \citep{van-Leeuwen2007a}.\\
}
\end{deluxetable*}

\subsection{MOSFIRE $Y$-band Observations in EGS}
Spectroscopic observations of our sample were obtained over two nights in April 2021 using the Keck MOSFIRE spectrograph \citep{McLean2012a}.  This observing time was awarded through the NASA/Keck allocation (PI: I. Jung).  We created four slitmask configurations that accommodate 61 high-redshift candidate galaxies for Ly$\alpha$ emission within a total effective sky area of $\sim10^\prime\times10^\prime$ (Figure \ref{fig:masks}).  We observed two pointings each night, resulting in $\sim$3.5 hr of total exposure time per mask. We used the $Y$-band filter to cover Ly$\alpha$ at $7.0<z<8.2$. The spectral resolution of the $Y$-band filter is $\sim3$\AA\ ($R=3500$), and the slit width was set to be 0\farcs7, which corresponds to the typical seeing level at Mauna Kea. During the observations, individual science frames were taken with 180-sec exposures, and we used a standard ABAB dither pattern (+1\farcs25, -1\farcs25, +1\farcs25, -1\farcs25). The seeing level varies through the nights from 0\farcs7 to 1\farcs2. The observational details are listed in Table \ref{tab:observations}.

\subsection{Data Reduction and Flux Calibration}
We used the recent version of the public MOSFIRE data reduction pipeline (DRP)\footnote{https://keck-datareductionpipelines.github.io/MosfireDRP/} to reduce the raw data. The public DRP provides a sky-subtracted, flat-fielded, and rectified two-dimensional (2D) slit spectrum per slit object. The reduced spectra are wavelength-calibrated using telluric sky emission lines.  Reduced 2D spectra have the spectral resolution of 1.09\AA\ pixel$^{-1}$ and the spatial resolution of 0\farcs18 pixel$^{-1}$. 

It has been reported that there is significant slit drift in the spatial direction (up to $\sim$pixel hr$^{-1}$) in previous MOSFIRE observations \citep[e.g.,][]{Kriek2015a, Song2016b, Jung2019a, Hutchison2020a, Larson2022a}, which needs to be handled separately if observations last longer than a couple of hours of exposure time. However, the general use of the public DRP is not aimed to correct the known slit drift in the spatial direction for long-exposure science. We corrected the slit drifts found in our observations, following \cite{Jung2020a}.  Briefly, we reduced each adjacent pair of science frames with the public DRP separately, generating the reduced 2D spectra of 360 sec exposure time. In our observations, we placed slits on two faint stars per slitmask for flux calibration and used them to trace the slit drifts in individual MOSFIRE pointings as well. The amount of slit drift is estimated by tracing the spatial positions of slit continuum sources on the DRP-reduced 2D spectra of 360-sec exposure. We corrected the measured slit drifts when combining 360-sec DRP-reduced 2D spectra to generate a single science frame for each slit target. Cosmic ray rejection and/or bad pixel cleaning are not feasible with the DRP runs on a pair of science frames. Thus, we cleaned them by taking sigma-clipped means in the 2D combination step.  Also, to maximize a resulting signal-to-noise ratio (SNR), we weight the DRP-reduced frames with the Gaussian peak fluxes of the slit stars, which reflect observing conditions. 

The one-dimensional (1D) spectra of our slit objects were extracted from the combined 2D spectra using an optimal extraction scheme \citep{Horne1986a} with a 1\farcs4 spatial window twice the typical seeing level of Mauna Kea. We model a spatial weight profile that follows the spatial profile of the slit stars, thus the pixels near the peak of the stellar spatial profile are maximally weighted. This enables us to correct the offsets of the actual spatial locations of slit objects from the expected positions, which are found up to a couple of pixels in a spatial direction. 

The reduced 1D spectra were used to search for emission-line candidates. For the detection candidates, we repeated 1D extraction by shifting the centers of the optimal 1D extraction within $\pm$3 pixels from the corrected spatial locations of our slit objects. This accounts for the uncertainties in the centering of the objects' spatial locations, allowing us to obtain maximum SNRs of the emission-line candidates.

For absolute flux calibration and telluric absorption correction, we used long-slit observations of a spectro-photometric standard star (HIP56147, the spectral type A0V) and \cite{Kurucz1993a} model stellar spectra. We estimated a wavelength-dependent response curve each night by dividing the model stellar spectrum with the reduced long-slit stellar spectra. The response curves were scaled to match the known photometric magnitude of the standard star.  However, our science observations were obtained in different observing conditions, such as seeing and airmass, to the standard star observations.  Thus, we estimated additional scaling factors using slit stars in the science slitmasks to refine the absolute flux calibration by matching their $Y$-band magnitudes measured from our spectra to the known $Y_{\text{105}}$ magnitudes from the existing HST photometry \citep{Finkelstein2022a}.  The additional scaling factors were at the level of $<10\%$.  The slit losses due to the narrow slit width of 0\farcs7 in our observations are corrected in this step, considering the seeing conditions. We assume our high-redshift target galaxies are point sources, as they are unresolved in our observations. Overall, we obtained 3$\sigma$ detection limits of emission lines at $\sim$5$\times10^{-18}$ erg s$^{-1}$ cm$^{-2}$ between sky-emission lines (with $\sim$3.5hr integration), and this is comparable to typical detection limits from previous MOSFIRE $Y$-band observations \citep[e.g.,][]{Finkelstein2013a, Song2016b, Jung2020a}.

\section{Results}
\subsection{Emission-Line Search}
We implemented an automated search scheme to capture plausible emission-line candidates consistently, similar to the method in \cite{Jung2020a}.  We first collected emission-line candidates that were selected via our automated search on both 1D and 2D spectra by performing Gaussian line fitting on the 1D and {\sc Source Extractor} \citep{Bertin1996a} runs on the 2D spectra. We required a 3$\sigma$ detection threshold in both the 1D and 2D searches. Then, we manually inspected individual emission-line candidates to rule out (i) sky-emission residuals, (ii) spurious sources, and (iii) contaminants from nearby sources. We conservatively removed emission-line features that are found close to the edge of sky-emission lines. To rule out spurious sources, we inspected the 2D spectra to ensure that there are clear negative peaks shown at the expected locations, $\pm$\,2\farcs5 apart from source positions, caused by the dither pattern of MOSFIRE. Additionally, we inspected the HST images to see if there are potential nearby contaminants whose emission lines could be captured at the same spatial locations of the MOSFIRE slits. Lastly, we performed tailored asymmetric (for the extended emission) or Gaussian (for the sharp/unresolved) emission-line fitting in reduced 1D spectra to calculate the line fluxes of emission-line candidates.
 
For emission-line-detected objects, we further checked their possibility of being low-redshift interlopers to ensure their nature as Ly$\alpha$. First, we checked if multiple emission lines are found in the same object. These would originate from a combination of emission lines from low-redshift galaxies (e.g., [\ion{O}{iii}] $\lambda\lambda$4959, 5007; H$\beta$; [\ion{N}{ii}] $\lambda\lambda$6548, 6584; H$\alpha$).  We manually inspected the wavelengths of the possible companion lines, and we find no evidence of multiple emission lines in our sample.  Second, we checked the possible low-redshift solution of being an [\ion{O}{ii}] $\lambda\lambda$3727, 3729 emitter which can mimic the Lyman-break feature with the Balmer break of low-redshift galaxies. If that is the case, the [\ion{O}{ii}] doublet should be resolved with the spectral resolution of Keck/MOSFIRE ($R=3500$ or $\sim$3\AA). However, none of our emission-line candidates display the doublet emission lines with a $\sim$7--8\AA\ separation (an expected peak separation of the [\ion{O}{ii}] doublet at $z\sim$ 1.7--1.8).  

\begin{figure*}[th]
\centering
\includegraphics[width=0.82\paperwidth]{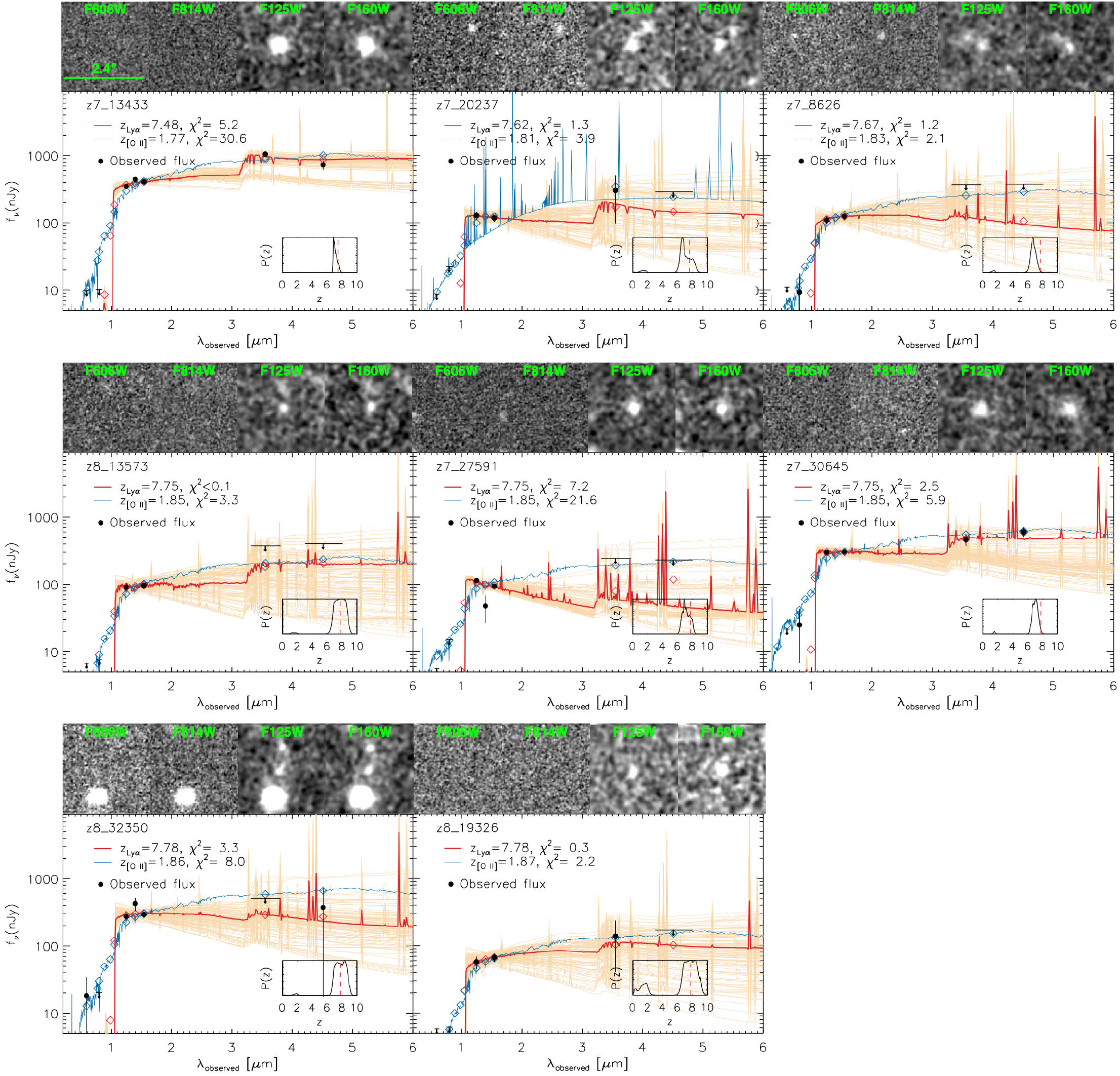}
\caption{The best-fit model SEDs of our $z>7$ LAEs. The best-fit model SEDs of high-redshift solutions (Ly$\alpha$) are shown as red curves, and the 100 random draws are displayed as thin lines in each panel. We also show the best-fit SEDs from low-redshift solutions of [\ion{O}{ii}] as blue curves for comparison. The photometric data in HST and \textit{Spitzer}/IRAC filters are shown as black symbols, and the downward arrows represent $1\sigma$ upper limits. The inset figures show the photometric redshift PDFs with spectroscopic redshifts from Ly$\alpha$ as dashed vertical lines.  The small HST cutouts of individual objects in the $V_{606}$, $I_{814}$, $J_{125}$, and $H_{160}$ filters are on top of each panel, which highlight strong Lyman-break between $I_{814}$ and $J_{125}$ images.}
\label{fig:seds}
\end{figure*}

We also compared the $\chi^2$ values from best-fit SEDs between high-redshift (with Ly$\alpha$) and low-redshift (with [\ion{O}{ii}]) solutions for Ly$\alpha$-emission candidates as supplementary check, removing Ly$\alpha$-emission candidates disfavored in SED fitting analysis. The best-fit model SEDs of the host galaxies are shown in Figure \ref{fig:seds}.  Our SED fitting analysis presents that the high-redshift solutions with Ly$\alpha$ are preferred over the low-redshift solutions for our Ly$\alpha$-detected galaxies in agreement with the photometric redshift PDFs.  

\begin{figure*}[th]
\centering
\includegraphics[width=0.82\paperwidth]{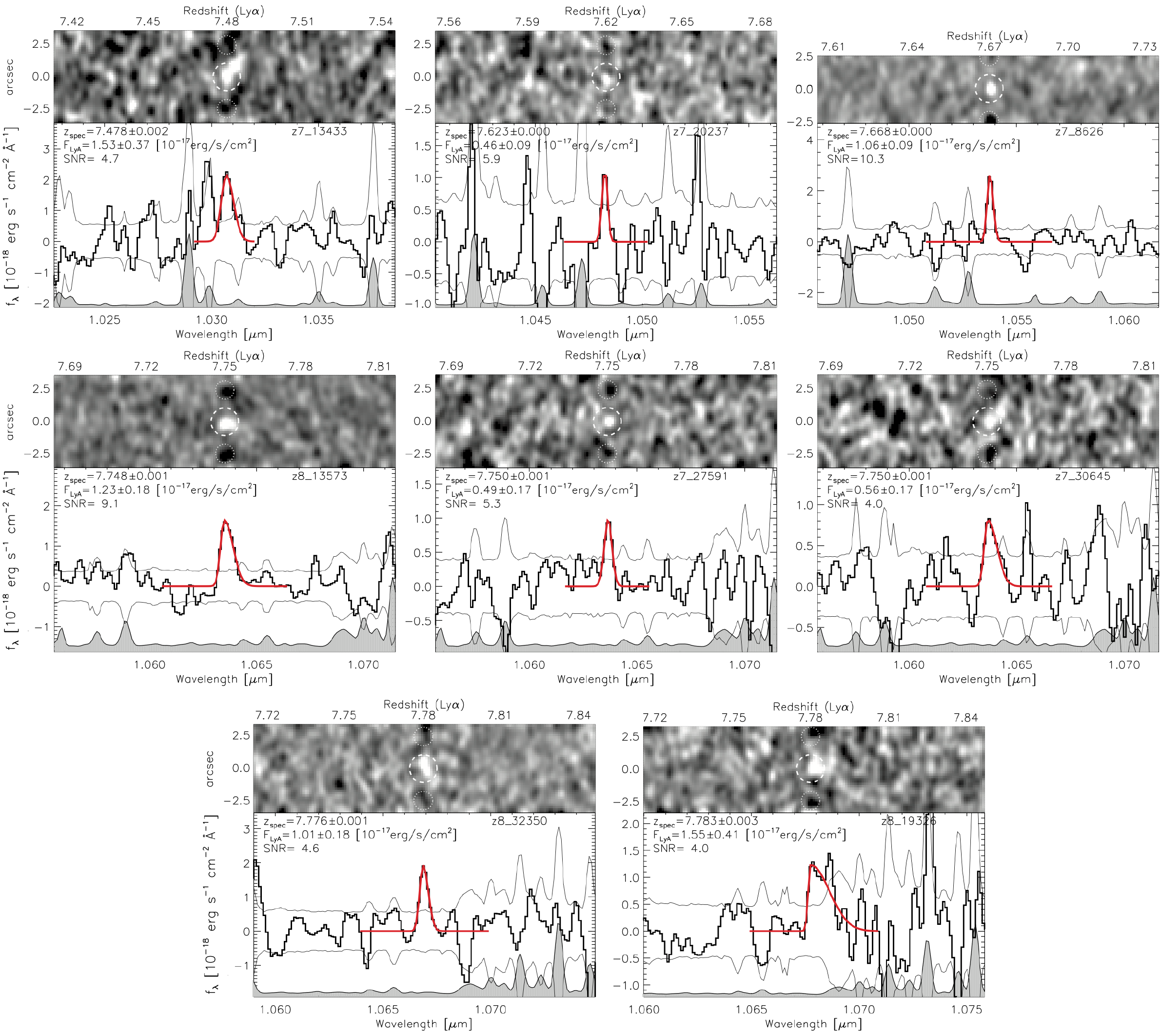}
\caption{MOSFIRE spectra of the detected Ly$\alpha$ emission lines: 2D on top and 1D at the bottom in each panel. In each panel, 1D and 2D spectra are centered at the detected emission line. At the bottom, the solid black lines are the 1D signals with 1$\sigma$ upper and lower bounds shown as thin grey curves.  Sky-emission regions are shown as the shaded curves at the bottom of each plot.  In 2D spectra, red circles denote the emission lines, and the negative traces caused by a dither pattern are marked with white circles. The red curves in 1D represent the best-fit asymmetric Gaussian curves.  All emission lines are found within the expected spatial locations in the 2D spectra (within $\pm3$ pixels in the y-axis), and two negative features, caused by our dither pattern, are shown at around $\pm$2\farcs5 from the detected emission lines.}
\label{fig:laes}
\end{figure*}

\begin{deluxetable*}{ccccccccccc}
\setlength{\tabcolsep}{0.03in}
\tablewidth{0pc}
\tabletypesize{\footnotesize}
\tablecaption{Summary of New Ly$\alpha$-Emission-Line and Host-Galaxy Properties\tablenotemark{a} \label{tab:LAE}} 
\tablehead{
\colhead{ID} & \colhead{R.A. (J2000.0)} &\colhead{Decl. (J2000.0)} & \colhead{$F_{\text{Ly}\alpha}$} & {SNR$_\text{2D}$\tablenotemark{a}}   & \colhead{EW$_{\text{Ly}\alpha}$\tablenotemark{b}} & \colhead{$z_{\text{Ly}\alpha}$}  & \colhead{$M_{\text{UV}}$} &  \colhead{$\beta$\tablenotemark{c}}  \\
	\colhead{}  & \colhead{degree}  & \colhead{degree}  & \colhead{(10$^{-17}$ erg s$^{-1}$ cm$^{-2}$)}  & \colhead{} & \colhead{(\AA)} & \colhead{} &\colhead{}  & \colhead{} \\
	\colhead{(1)}  & \colhead{(2)} & \colhead{(3)} & \colhead{(4)} & \colhead{(5)} & \colhead{(6)} & \colhead{(7)} & \colhead{(8)}  & \colhead{(9)}
}
\startdata
{z7\_13433} &  {214.85083} & {52.77666} & {1.53$\pm$0.37} & {4.7} & {22.2$^{+8.6}_{-7.0}$} &{7.4784$\pm$0.0019}  &{-22.1} & {-1.52$^{+0.24}_{-0.22}$}  \\  
{z7\_20237} &  {215.10658} & {52.97582} & {0.46$\pm$0.09} & {5.9} & {17.1$^{+8.6}_{-5.7}$}  &{7.6228$\pm$0.0003} &{-21.1} & {-2.23$^{+0.52}_{-0.49}$} \\ 
{z7\_8626} &  {215.11446} & {52.95123} & {1.06$\pm$0.09} & {10.3}  & {49.4$^{+17.5}_{-11.7}$} &{7.6682$\pm$0.0002} &{-21.0} & {-1.67$^{+0.44}_{-0.61}$}   \\  
{z8\_13573} &  { 215.15088} & {52.98957} & {1.23$\pm$0.18} & {9.1}  & {69.1$^{+29.8}_{-19.9}$} &{7.7482$\pm$0.0009} &{-20.7} & {-1.80$^{+0.47}_{-0.53}$} \\  
{z7\_27591} &  {215.13288} & {53.04786} & {0.49$\pm$0.17} & {5.3}  & {19.1$^{+9.0}_{-7.6}$}  &{7.7496$\pm$0.0007} &{-20.8}  & {-2.80$^{+0.39}_{-0.25}$} \\  
{z7\_30645} &  {215.09504} & {53.01421} & {0.56$\pm$0.17} & {4.0} & {8.7$^{+4.3}_{-3.4}$}  &{7.7496$\pm$0.0009}  &{-22.1} & {-2.18$^{+0.28}_{-0.33}$} \\  
{z8\_32350} &  {214.99903} & {52.94197} & {1.01$\pm$0.18} & {4.6}  & {17.7$^{+8.6}_{-5.7}$} &{7.7759$\pm$0.0012} &{-21.9} & {-1.98$^{+0.37}_{-0.55}$}  \\  
{z8\_19326} &  {215.11962} & {52.98284} & {1.55$\pm$0.41} & {4.0} & {151.0$^{+125.4}_{-66.2}$} &{7.7832$\pm$0.0035} &{-20.1} & {-1.56$^{+0.37}_{-0.62}$}   
\enddata
\tablecomments{Columns: (1) Object ID, (2) Right ascension, (3) Declination, (4) Ly$\alpha$ emission line flux, (5) Emission-line detection significance, (6) Rest-frame equivalent width of Ly$\alpha$ emission line, (7) spectroscopic redshift based on Ly$\alpha$ emission line, (8) galaxy UV magnitude estimated from the averaged flux over a 1450 -- 1550\AA\ bandpass from the best-fit galaxy SED model, (9) Rest-UV continuum slope.}
\tablenotetext{}{
$^{a}$\footnotesize Detection significance measured from {\sc Source Extractor} runs on 2D spectra.\\
$^{b}$Listed uncertainties account for the UV continuum measurement errors from SED fitting.\\
$^{c}$The power-law slope of the rest-frame UV continuum, measured from the best-fit SEDs. 
}
\end{deluxetable*}

From our emission-line search, we find eight galaxies with Ly$\alpha$ emission detected in the spectra (SNR $\geq$ 4). The 1D and 2D spectra of the detected emission lines are displayed in Figure \ref{fig:laes}. We caution that we are unable to completely rule out the chance of being low-redshift interlopers. The secondary lines that might have served to reject a high-redshift interpretation could be below the detection limit, particularly when coincident with a strong sky emission line. The strongly varying signal-to-noise can be seen in Figure \ref{fig:laes}. Nevertheless, based on our robust identification of emission lines and thorough diagnosis of low-redshift interlopers, we conclude that the detected emission lines are consistent with Ly$\alpha$ emission at $z>7$.  

\begin{figure*}[ht]
\centering
\includegraphics[width=0.80\paperwidth]{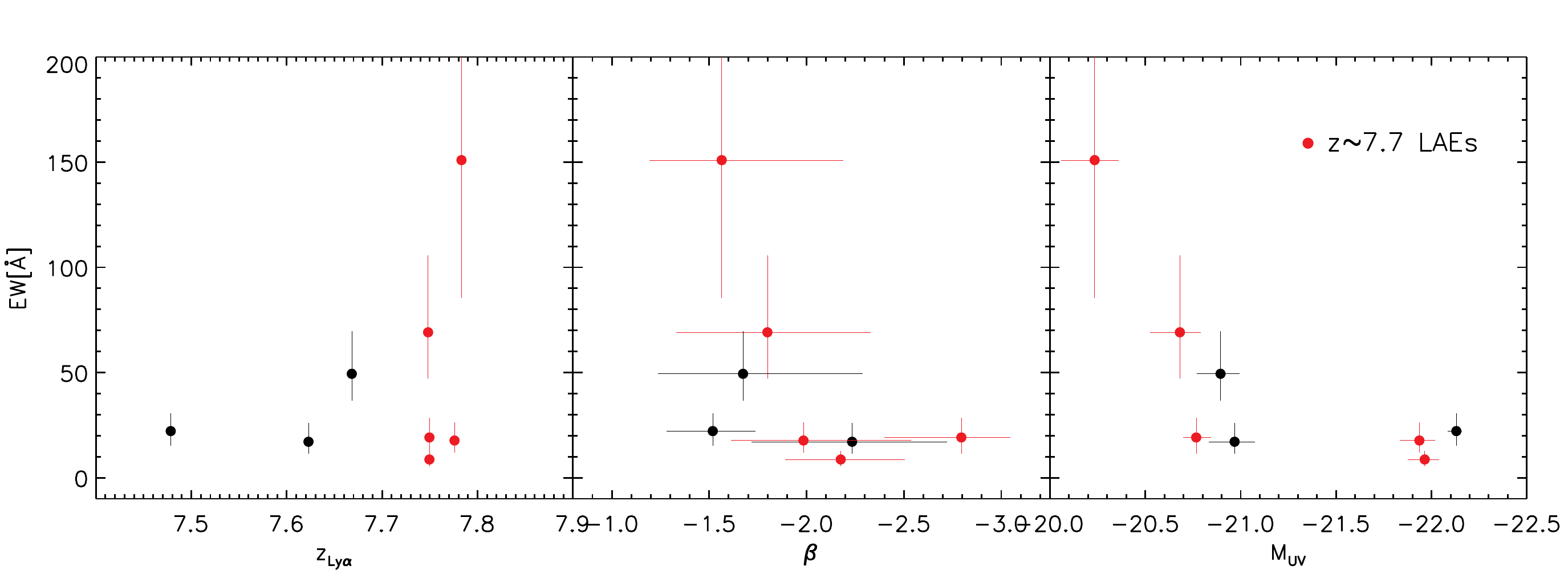}
\caption{Rest-frame Ly$\alpha$ EWs versus redshift (left), the rest-UV continuum slope (middle), and $M_\text{UV}$ (right).  The $z\sim7.7$ LAEs are symbolized with red circles. Interestingly, two of the $z\sim$ 7.7 LAEs, which are the faintest in UV among our LAEs, emit the highest EW Ly$\alpha$ emission lines (EW $>50$\AA) as shown in the right panel.}
\label{fig:ewplot}
\end{figure*}

\subsection{$z>7$ Ly$\alpha$ Emitters} 
\subsubsection{Emission-Line Properties}
Table \ref{tab:LAE} summarizes the properties of the detected Ly$\alpha$-emission lines and their host galaxies. To measure emission-line properties, we performed 1D (asymmetric) Gaussian fitting to the reduced 1D spectra. The fiducial values were taken from the best-fit (asymmetric) Gaussian curves. To estimate the 1$\sigma$ errors of the emission-line properties, we perform the same (asymmetric) Gaussian fitting to 1000 Monte Carlo realizations of the perturbed 1D spectra with corresponding error spectra. The spectroscopic redshifts are calculated from the peak wavelength of the best-fit Gaussian curves, and the line fluxes are obtained from the total fluxes under the Gaussian curves. The rest-frame equivalent width (EW) is estimated as:
\begin{eqnarray}
EW = \frac{F_{\text{Ly}\alpha}}{f_{\text{cont}}(1+z_{\text{Ly$\alpha$}})}, 
\end{eqnarray} 
 where $F_{\text{Ly$\alpha$}}$ is the Ly$\alpha$ emission-line flux, and $f_{\text{cont}}$ is the continuum flux density, derived by averaging rest-UV continuum of the best-fit SEDs within a 50\AA-wavelength window of the rest-frame 1230 -- 1280\AA.

In Figure \ref{fig:ewplot}, we show the Ly$\alpha$ EWs versus redshift (left), the rest-UV slope (middle), and $M_{\text{UV}}$ (right). In the left panel, we highlight that five of our eight LAEs are clustered at $z\sim7.7$ (red symbols) in close proximity to the known three LAEs at $z\sim7.7$ presented in \cite{Tilvi2020a}. Although there is no significant trend seen with the rest-UV slope ($\beta$) in the middle panel, our LAEs are showing rest-UV slopes mostly bluer than $\beta\lesssim-1.5$. Interestingly, two of them emit the highest EW Ly$\alpha$ ($EW>50$\AA) that are faintest in UV among our LAEs (in the right panel).  We will discuss more details on the clustered LAEs in Section 4.

\section{Extended Ionized Structure around Clustered LAEs at $z\sim7.7$}
\subsection{Clustered LAEs at $z\sim7.7$}
The process of reionization is expected to be inhomogeneous as overdensity regions with clustered galaxies were to be ionized earlier than field areas \citep[e.g.,][]{Finlator2009a, Mesinger2015a, Katz2019a}.  The currently most accessible tool to probe ionized structures in the middle of reionization is to search for Ly$\alpha$ emission from the reionization-era galaxies. A recent effort to spectroscopic search for Ly$\alpha$ resulted in the discoveries of clustered LAEs in the middle/late phase of reionization at $z\sim$ 7 -- 8 \citep[e.g.,][]{Zheng2017a, Castellano2018a, Tilvi2020a, Jung2020a, Endsley2021b}. Particularly, the EGS field has a couple of known clustered structures with LAEs at $z\gtrsim7.5$ \citep{Zitrin2015a, Oesch2015a, Roberts-Borsani2016a, Tilvi2020a, Larson2022a}, and their additional membership candidates are discussed based on photometric selection \citep{Leonova2022a}. Our MOSFIRE $Y$-band program detected Ly$\alpha$ emission from eight sources, and five of them are potentially associated with the known $z\sim7.7$ LAE cluster \citep{Tilvi2020a}\footnote{Ly$\alpha$ from the brightest galaxy (z8\_5) was first detected in \cite{Oesch2015a} and \cite{Roberts-Borsani2016a}, and its \ion{C}{iii}] emission was detected in \cite{Stark2017a}.}. This demonstrates a possible extension of the LAE structure with up to eight LAEs at $z\sim7.7$. This is currently the largest measured LAE cluster system in this early Universe at $z>7$. We summarize the $z\sim7.7$ LAEs in Table \ref{tab:clusterLAEs}.

\subsection{Extended Ionized Structure}
\begin{figure*}[t]
\centering
\includegraphics[width=0.5\paperwidth]{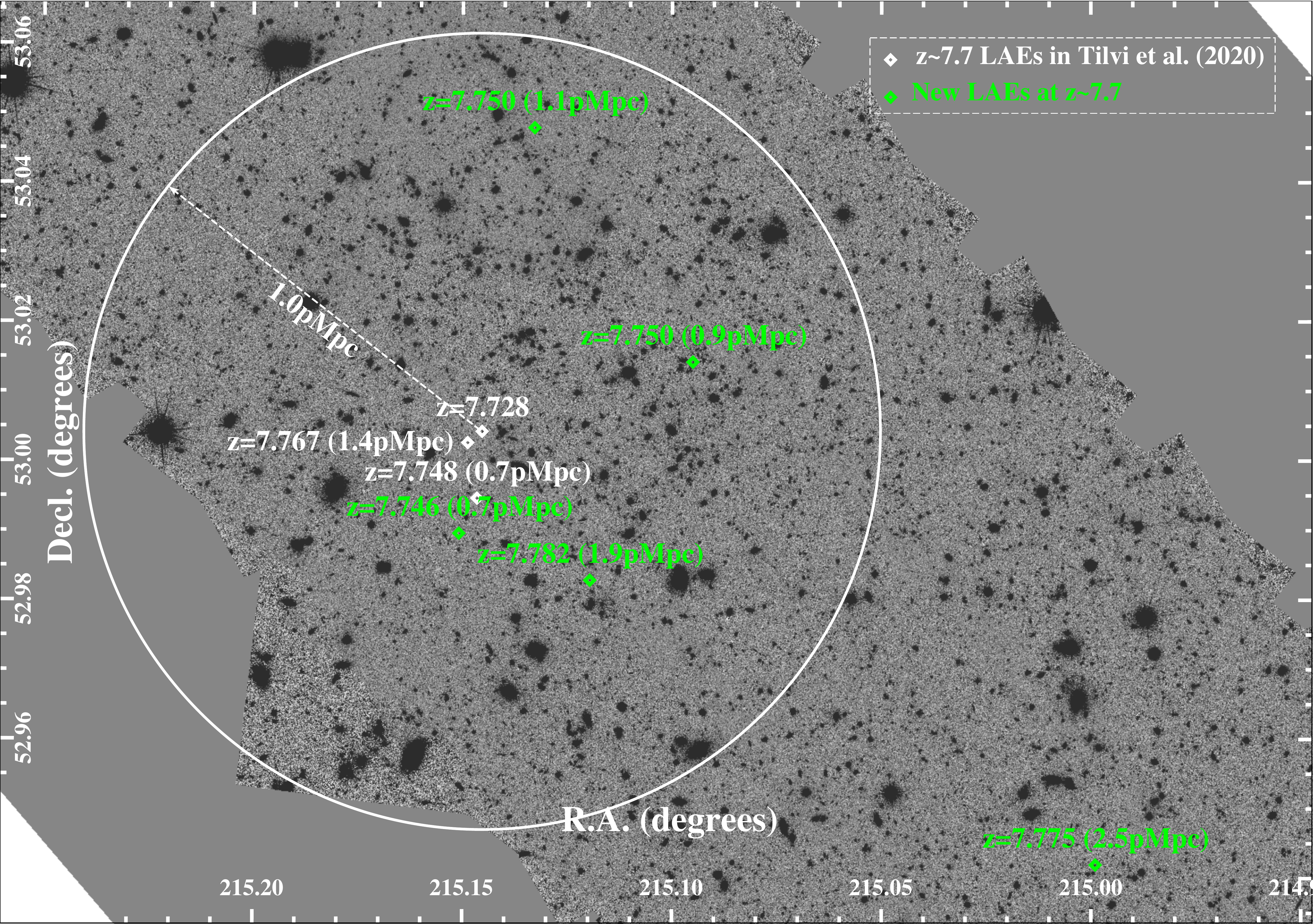}
\includegraphics[width=0.33\paperwidth]{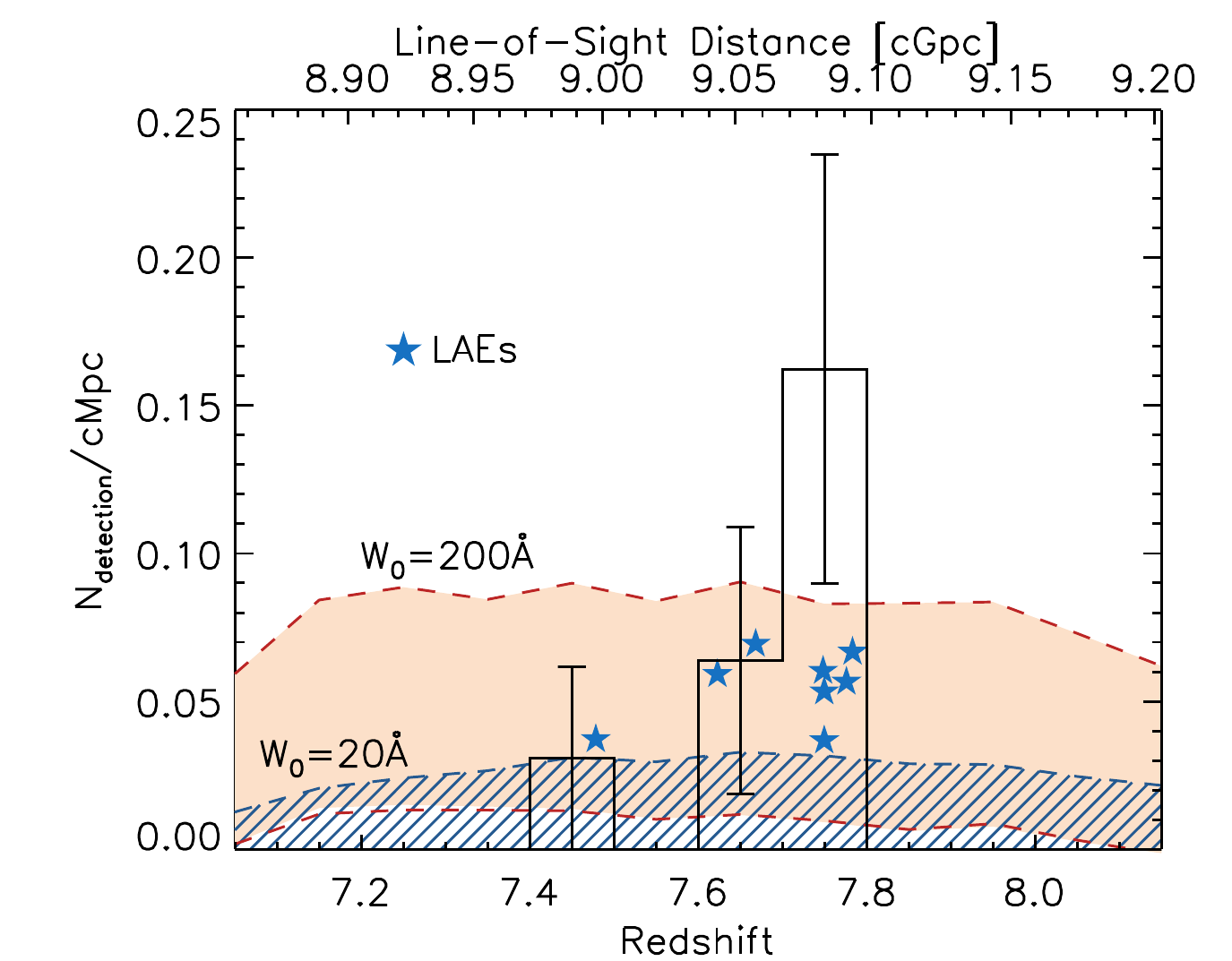}
\caption{(Left) Spatial distribution of $z\sim7.7$ LAEs in the CANDELS EGS field, displayed on the HST \textit{H}-band image. Three $z\sim7.7$ LAEs reported in \cite{Tilvi2020a} are shown as white diamond points, and the new $z\sim7.7$ LAEs discovered in this work are marked as green symbols. The white circle represents a 1pMpc-radius ionized bubble, which is estimated to be produced by the brightest galaxy (z8\_5) in \cite{Tilvi2020a}, and the 3D spatial separations of new $z\sim7.7$ LAEs from z8\_5 are shown in parentheses.  Seven $z\sim7.7$ LAEs are clustered within $\lesssim$2pMpc, and an additional LAE is at a 2.5pMpc distance. As the estimation of ionized bubble size is typically based on a single source, an extended ($>$2.5pMpc) ionized structure could be created by overlapping ionized bubbles. (Right) LAE density as a function of redshift. The plot shows the number of Ly$\alpha$ detections above observational detection limits per unit volume, a 1cMpc-think slice in the line-of-sight (LOS) direction. The actual detections from our observations are shown as histograms with the Poissonian errors in individual bins.  The blue star symbols denote spectroscopic redshifts of the LAEs, that are arbitrarily distributed along the y-axis to avoid overlaps.  The shaded regions represent the 1$\sigma$ ranges of the expected numbers of Ly$\alpha$ detections from our EW distribution modeling. The case of the high Ly$\alpha$ EW distribution with $W_0=200$\AA\ is shown as a red-shaded area, and the low Ly$\alpha$ EW case of $W_0=20$\AA\ is displayed with a diagonal-pattern filled region. This plot reveals a clear spike at $z\sim7.7$. The $z\sim7.7$ spike is more comparable to the extreme case of the high Ly$\alpha$ EW distribution whereas the low Ly$\alpha$ EW case explains well the low/non-detections of Ly$\alpha$ at other redshift ranges.}
\label{fig:laecluster}
\end{figure*}

With the discovery of the clustered LAEs at $z\sim7.7$, we calculate 3-dimensional (3D) separations of individual $z\sim7.7$ LAEs from the brightest (and potentially central) galaxy (z8\_5). The left panel of Figure \ref{fig:laecluster} shows the distribution of our $z\sim7.7$ LAEs (green) in addition to the three $z\sim7.7$ LAEs that were previously discovered \citep[white;][]{Oesch2015a, Roberts-Borsani2016a, Tilvi2020a}. The estimated 3D physical distances from z8\_5 are listed in parentheses, ranging from 0.7pMpc at the nearest to 2.5pMpc at the farthest. 

The crowd of ionizing sources could create the extended ionized structure beyond a $\sim$1pMpc scale of individual ionized bubbles and eventually enhance Ly$\alpha$ transmission in the IGM.  To examine whether these clustered objects are situated in a connected ionized structure, we estimated the ionized bubble sizes which could be created by individual LAEs.  Following \cite{Tilvi2020a} and \cite{Jung2020a}, we used the relation between Ly$\alpha$ luminosities and ionized bubble sizes, predicted in the theoretical models from \cite{Yajima2018a}.  Briefly, \cite{Yajima2018a} model LAEs and the ionized bubble sizes based on individual halo merger trees using star formation history which is modeled to provide a reionization history consistent with the Planck observations \citep{Planck-Collaboration2016a}.  Based on the \cite{Yajima2018a} models, we used the measured Ly$\alpha$ luminosities to derive the predicted sizes of individual ionized bubbles around LAEs. The estimated ionized bubble sizes are ranging from $\sim$0.7 to 1.0 pMpc, as listed in Table \ref{tab:clusterLAEs}.  As the models in \cite{Yajima2018a} predict the growth of isolated ionized bubbles around LAEs, it does not consider the additional expansion due to the overlapping ionized bubbles.  Thus, the derived ionized bubble sizes may indicate the lower limits of ionized bubble sizes around LAEs; a much larger ionized structure could be created by the overlaps of multiple ionized bubbles around these LAEs.  Such an extended ionized structure may promote Ly$\alpha$ escape from galaxies \citep[e.g.,][]{Mason2020a, Park2021a, Qin2021a, Smith2021a}, resulting in enhanced Ly$\alpha$ detection rate in our observations at this redshift.

\begin{deluxetable*}{cccccccc}
\tabletypesize{\footnotesize}
\tablecaption{Summary of Ly$\alpha$ Emitters at the $z\sim7.7$ Overdensity \label{tab:clusterLAEs}} 
\tablehead{
\colhead{ID} & \colhead{R.A. (J2000.0)} &\colhead{Decl. (J2000.0)} & \colhead{$z_{\text{Ly}\alpha}$} & \colhead{$L_{\text{Ly}\alpha}$} & \colhead{\ion{H}{II} radii} & {$\Delta_\text{3D}$ from z8\_5}  &\colhead{$M_\text{UV}$} \\
	\colhead{}  & \colhead{degree}  & \colhead{degree}  & \colhead{} &\colhead{(10$^{43}$ erg s$^{-1}$)}  & \colhead{(pMpc)}  & \colhead{(pMpc)} &\colhead{}  \\
	\colhead{(1)}  & \colhead{(2)} & \colhead{(3)} & \colhead{(4)} & \colhead{(5)} & \colhead{(6)} & \colhead{(7)} & \colhead{(8)} }
\startdata
\multicolumn{8}{c}{\textbf{New Ly$\alpha$ Emitters in this work}} \\
{z8\_13573} &  {215.15088} & {52.98957} &{7.7482$\pm$0.0009}&{$0.93 \pm 0.14$} &{0.93} &{0.7}  &{-20.7}\\  
{z7\_27591} &  {215.13288} & {53.04786} &{7.7496$\pm$0.0007} &{$0.37 \pm0.13$} &{0.68}  & {1.1} & {-20.8} \\  
{z7\_30645} &  {215.09504} & {53.01421} &{7.7496$\pm$0.0009}  &{$0.42 \pm0.13$} &{0.71}  &{0.9}	  &{-22.1} \\  
{z8\_32350} &  {214.99903} & {52.94197} &{7.7759$\pm$0.0012} &{$0.77 \pm 0.14$} &{0.87}  &{2.5}	 &{-21.9}   \\  
{z8\_19326} &  {215.11962} & {52.98284} &{7.7832$\pm$0.0035} &{$1.18 \pm0.31$} &{1.01} &{1.9}  &{-20.1}  \\ 
\hline
\multicolumn{8}{c}{\textbf{Ly$\alpha$ Emitters in \cite{Tilvi2020a}}\tablenotemark{a}} \\
{z8\_5} &  {215.14530} & {53.00423} &{7.728} & {1.2$\pm$0.1} & {1.02}  &{-}& {-22.3\tablenotemark{b}} \\
{z8\_4} &  {215.14654} &  {52.99461} &{7.748} & {0.4$\pm$0.1} & {0.69}  &{0.7}& {$>$-20.3} \\
{z8\_SM} & {215.14873} & {53.00259} &{7.767} & {0.2$\pm$0.1} & {0.55}  &{1.4}& {$>$-20.3} \\
\enddata
\tablecomments{Columns: (1) Object ID, (2) Right ascension, (3) Declination, (4) spectroscopic redshift based on Ly$\alpha$ emission line, (5) Ly$\alpha$ emission luminosity, (6) radii of ionized \ion{H}{II} bubble around LAEs based on the relation between Ly$\alpha$ luminosities and the bubble sizes from the \cite{Yajima2018a} model (see more discussion in Section 4.2), (7) Physical 3D separation from z8\_5,(8) galaxy UV magnitude estimated from the averaged flux over a 1450 -- 1550\AA\ bandpass from the best-fit galaxy SED model.}
\tablenotetext{}{
$^\text{a}$ The listed values are taken from \cite{Tilvi2020a}.\\
$^\text{b}$ $M_\text{UV}$ for this object is not given in \cite{Tilvi2020a}, thus we calculate it from our SED fitting analysis as same as done for other galaxies.
}
\end{deluxetable*}

\subsection{Enhanced Ly$\alpha$ Detection Rate at $z\sim7.7$}
The detectability of Ly$\alpha$ emission in targeted spectroscopic observations is affected by target selection functions (which considers photometric redshift measurement PDFs and galaxy $M_{UV}$ distribution) and detection limits (depending on e.g., observing conditions and the presence of sky-emission lines) in addition to the Ly$\alpha$ IGM transmission particularly during the epoch of reionization. Thus, this makes it complicated to interpret a Ly$\alpha$ detection rate at its face value.  

Instead, we performed Ly$\alpha$ EW distribution modeling to estimate the expected number of Ly$\alpha$ detections in our observations, which also consider target selection as well as observational conditions. Following \cite{Jung2020a}, in our Ly$\alpha$ EW modeling, we assume the Ly$\alpha$ EW distribution in its exponential functional form, $dN/dEW\propto \text{exp(-EW)}/W_0$, characterized with an $e$-folding scale ($W_0$).  We populate mock Ly$\alpha$ emission lines for spectroscopic targets with (i) EW values that are randomly taken from the assumed EW distributions and (ii) wavelength locations, also randomly chosen based on galaxy photometric redshift PDFs. We calculate the expected detection rates above the detection limits of these simulated Ly$\alpha$ emission lines.  To sum up, we quantify the Ly$\alpha$ detectability into the expected number of detections above detection limits as a function of $W_0$ via our EW modeling.

In the right panel of Figure \ref{fig:laecluster}, we compare the actual Ly$\alpha$ detection to what is estimated in our EW modeling.  The figure shows the number of Ly$\alpha$ detection above detection limits per unit volume, a 1cMpc-think slice in the line-of-sight (LOS) direction in the sky area covered in our observations.  We present the 1$\sigma$ range of the expected Ly$\alpha$ detections for the high and low Ly$\alpha$ EW cases with $W_0=200$\AA\ and 20\AA, respectively.  The choice of $W_0=200$\AA\ represents the distribution of extremely large EW Ly$\alpha$ emission lines whereas the low Ly$\alpha$ EW case of $W_0=20$\AA\ is comparable to the statistical measurement of $W_0$ from $M_\text{UV}<-20$ galaxies in this redshift \citep{Jung2022a}.  In the figure, the redshift distribution of our actual Ly$\alpha$ detections are shown as the blue star symbols, and the black histogram shows the estimated detection number density per unit volume.  The spike of our actual Ly$\alpha$ detection at $z\sim7.7$ exceeds the expectation of the extreme case of the high Ly$\alpha$ EW distribution ($W_0=200$\AA\; red shades) whereas non/rare detections of Ly$\alpha$ at other redshift ranges are more consistent with the low Ly$\alpha$ EW case ($W_0=20$\AA).  Even without the three known LAEs of \cite{Tilvi2020a}, our Ly$\alpha$-detection-rate analysis demonstrates that we observe significantly stronger Ly$\alpha$ from the clustered galaxies compared to that from the rest of the galaxies. 

\begin{figure*}[ht]
\centering
\includegraphics[width=0.58\paperwidth]{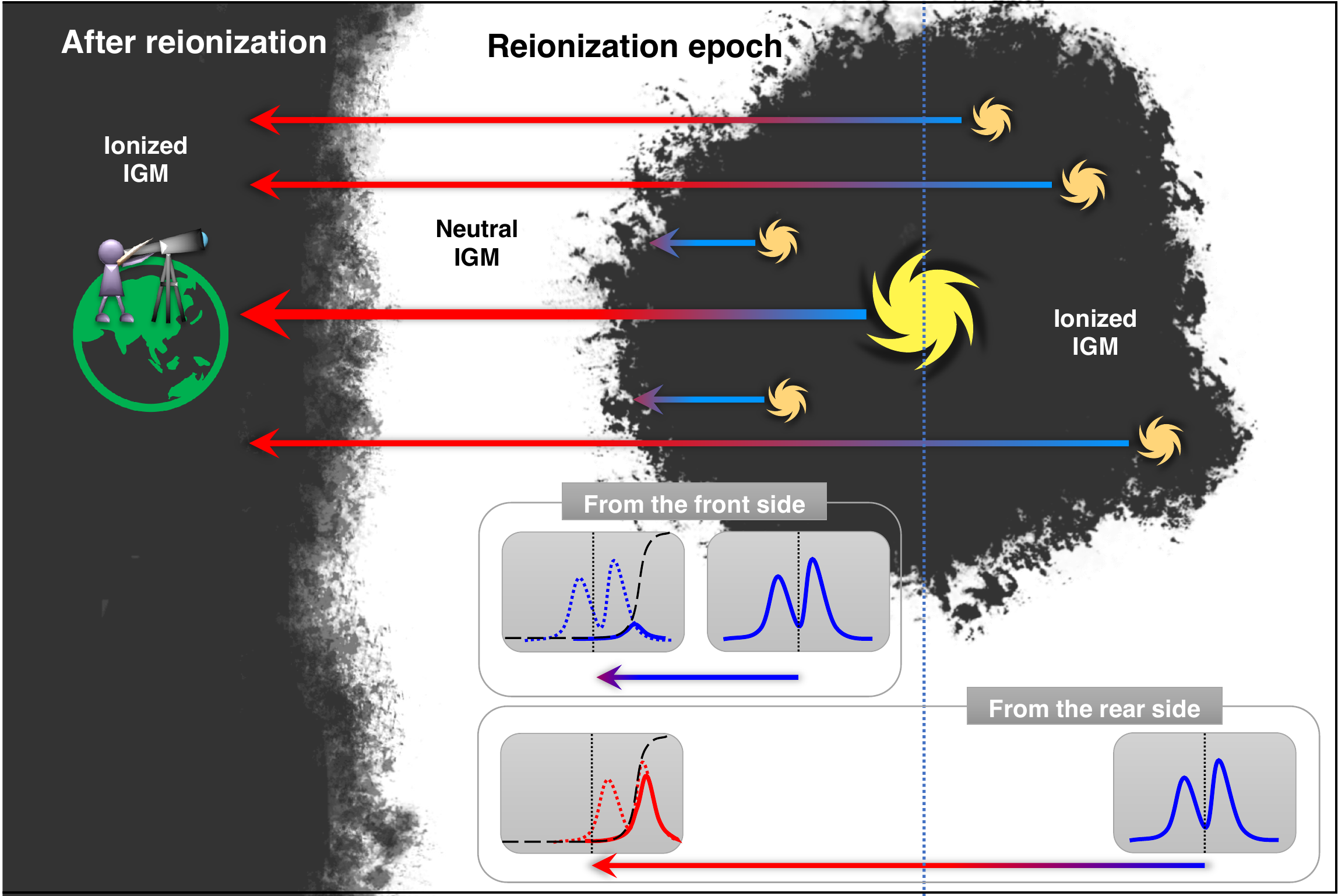}
\includegraphics[width=0.25\paperwidth]{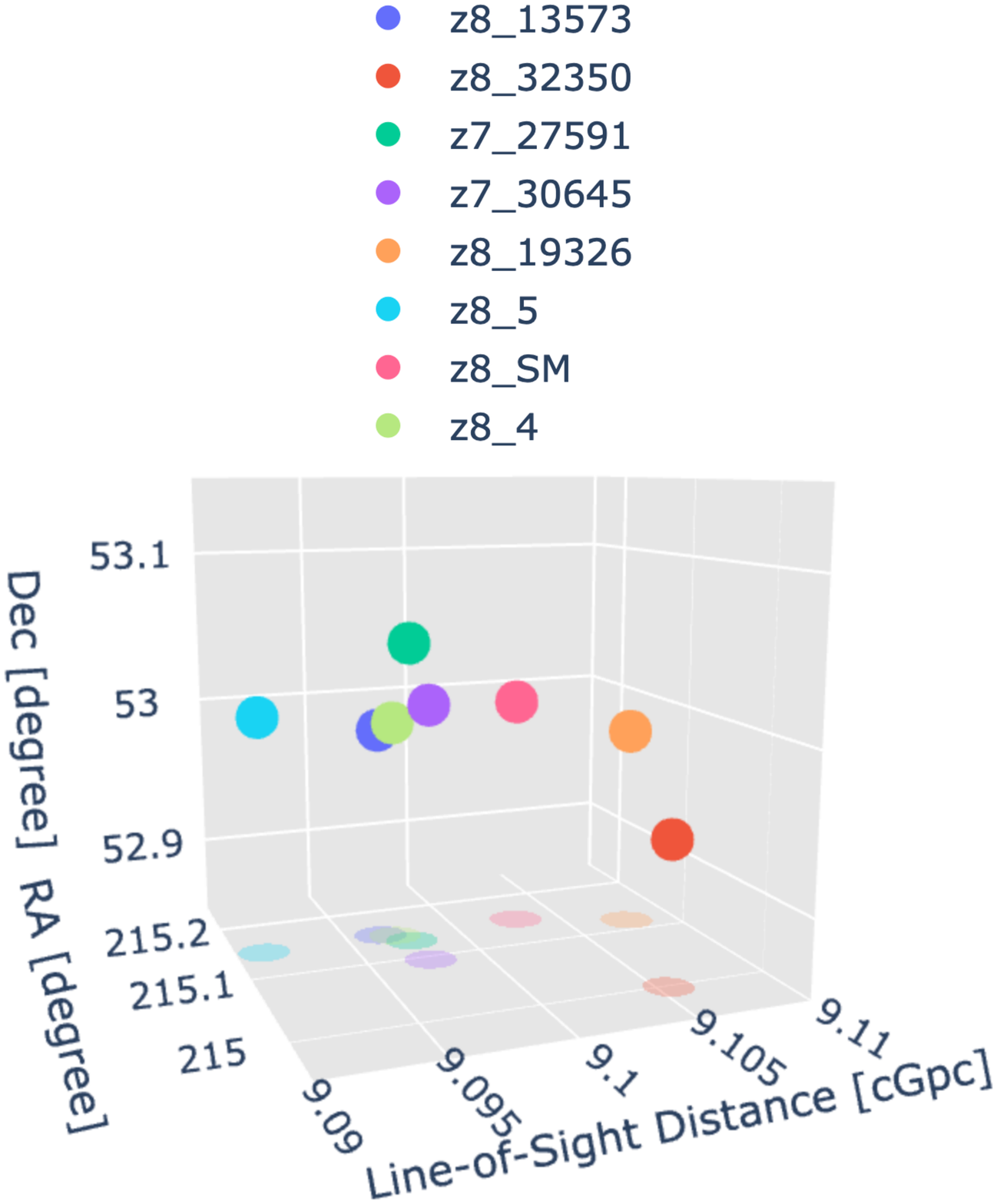}
\caption{(Left) Cartoon depicting boosted Ly$\alpha$ visibility through the IGM from galaxies in the rear side of ionized bubbles. The inset figures at the bottom illustrate the Ly$\alpha$ line shift in the rest frame of the neutral IGM.  The black long-dashed lines in the left inset panels show the wavelength-dependent IGM transmission. 
Ly$\alpha$ from the rear-side galaxies is significantly shifted to red from the central velocity while the line shift is much less for Ly$\alpha$ from the front-side galaxies.
This results in relatively boosted IGM transmission to Ly$\alpha$ from galaxies in the rear side of ionized bubbles. (Right) Spatial distribution of the clustered LAEs, including three LAEs of \cite{Tilvi2020a}: z8\_5, z8\_4, and z8\_SM.  The faint symbols represent the projections in the bottom plane. The brightest galaxy (z8\_5) is shown as the cyan point, which is placed in front of the other LAEs in the LOS direction.}
\label{fig:cartoon}
\end{figure*}

\subsection{Boosted IGM Transmission of Ly$\alpha$ from Galaxies in the Rear Side of Ionized Bubbles}
As discussed above, Ly$\alpha$ transmission can be enhanced in an extended ionization structure formed by clustered galaxies because the Ly$\alpha$ transmissivity increases with the distance between the source and the boundary of the ionized region \citep[e.g.,][]{Dijkstra2014b}. 
It is remarkable that the brightest member (z8\_5) of the clustered LAEs has the lowest redshift in the cluster indicating that it is likely located in front of all the other LAEs toward the observer (see the right panel of Figure \ref{fig:cartoon}). That is, we may be witnessing a bright galaxy boosting the Ly$\alpha$ visibility preferentially on its rear side. 

The left panel of Figure \ref{fig:cartoon} illustrates how Ly$\alpha$ transmissivity can be boosted in the rear side of a bright galaxy. The bright galaxy may be located at the center of the local ionized bubble. The ionized bubble opens a wavelength window on the red side of Ly$\alpha$ where photons can avoid being resonantly scattered by the neutral IGM. Ly$\alpha$ photons from galaxies on the rear side experience a larger cosmological redshift before reaching the neutral region and have a better chance of transmitting through the window. Indeed, a simulation study by \citet{Park2021a} reported that fainter galaxies tend to have a larger variation in Ly$\alpha$ transmission as their transmissivity sensitively depends on where they are located with respect to their brighter neighbors.

A peculiar motion of galaxies can contribute to the line shift as well. However, the redshift range of these clustered LAEs corresponds to $\gtrsim 1000$ km\,s$^{-1}$ in velocity, which is too large to be explained by the gravitational dynamics of galaxies alone. In theoretical studies, galaxies that are similar to z8\_5 in UV magnitude have their total masses around $\sim 10^{12}M_{\odot}$ \citep[e.g.,][]{Ocvirk2020a}. The gravitational infall velocity of such a galaxy peaks at $\sim$250--300 km\,s$^{-1}$ near the virial radius ($\sim$0.04 pMpc at $z=7.7$) and decreases as $r^{-0.5}$ with increasing distance to the galaxy ($r$).  As the clustered LAEs spread out over a $\gtrsim$1pMpc distance, we conclude that the LAEs observed here are far from forming a dynamically relaxed system in this early Universe.

\subsection{Comparison with UV-faint Galaxy Observations}
Recent surveys have produced a higher yield of Ly$\alpha$ detections from UV-luminous ($M_\text{UV}\lesssim-20$) galaxies than earlier attempts \citep[e.g.,][and in this work]{Castellano2018a, Jung2019a, Jung2020a}. In particular, spectroscopic observations of luminous sources with IRAC color excess, which reflects intense [\ion{O}{iii}]+H$\beta$ emission, delivered a much higher Ly$\alpha$ detection rate \citep[e.g.,][]{Oesch2015a, Zitrin2015a, Roberts-Borsani2016a, Stark2017a, Endsley2021a, Laporte2021a}.  In contrast, Ly$\alpha$ has rarely been detected in follow-up spectroscopic observations of UV-faint ($M_\text{UV}\gtrsim-20$) galaxies \citep{Hoag2019a, Mason2019a}.  Moreover, the unusually high detection rate of Ly$\alpha$ from galaxies with the IRAC excess is again somewhat diminished when it comes to targeting faint galaxies \citep{Roberts-Borsani2022a}.  Additionally, Ly$\alpha$ emission has not been detected in the recent JWST NIRSpec observations which targeted UV-faint galaxies \citep{Roberts-Borsani2022b, Williams2022a, Morishita2022a}.

This notable difference in a Ly$\alpha$ detection rate between UV-luminous and UV-faint populations could be understandable in the sense that an inhomogeneous nature of reionization suggests an earlier process of reionization around overdense regions where bright galaxies are preferentially located \citep[e.g.,][]{Finlator2009a, Katz2019a, Ocvirk2021a, Kannan2022a}.  Although \cite{Morishita2022a} report non-detection of Ly$\alpha$ even from clustered galaxies, those galaxies are less luminous (with $M_\text{UV}\gtrsim-20$) than those studied here.  Thus they may not create ionized regions large enough to allow the escape of Ly$\alpha$.

\section{Summary and Discussion}
We present our analysis of Keck/MOSFIRE $Y$-band spectroscopic observations for Ly$\alpha$ at 61 $7.0<z<8.2$ from 61 high-redshift candidate galaxies in the CANDELS EGS field, covering a total effective sky area of $\sim10^\prime\times10^\prime$. Most of our spectroscopic targets are relatively UV-bright ($M_\text{UV}\lesssim-20$). Our findings are summarized as follows.

\begin{enumerate}
\item We provide spectroscopic confirmations of Ly$\alpha$ ($>$4$\sigma$) from eight galaxies at $z>7$. This includes five potential members of the $z\sim7.7$ LAE cluster. Interestingly, two of them emit the highest EW Ly$\alpha$ emission lines (EW $>50$\AA) that are the faintest in UV among our LAEs.

\item The five $z\sim7.7$ LAEs from our observations are potentially associated with the known $z\sim7.7$ LAEs \citep{Tilvi2020a}, forming eight clustered LAEs at $z\sim7.7$. This is currently the largest measured LAE cluster system in this early Universe at $z>7$.

\item From our Ly$\alpha$ EW modeling, we estimated expected Ly$\alpha$ detection rates per unit volume in the line-of-sight (LOS) direction (or a redshift bin) depending on the choice of Ly$\alpha$ EW distribution. It suggests significantly stronger Ly$\alpha$ from the clustered $z\sim7.7$ LAEs, compared to the rest of our targets. 

\item We conclude that the clustered LAEs are likely to form an extended ionized structure around them based on the estimate of ionized bubble sizes around individual LAEs. The existence of such an extended ionized structure may allow the easier escape of Ly$\alpha$ from galaxies inside. This is aligned with the enhanced detection rate of Ly$\alpha$ at $z\sim7.7$.

\item We notice that the brightest object (z8\_5) in the $z\sim7.7$ LAE cluster is located at slightly lower (Ly$\alpha$) redshift than the other $z\sim7.7$ LAEs. This may indicate that we are witnessing the boosted IGM transmission of Ly$\alpha$ from galaxies that are situated on the rear side of an ionized area.

\item Our observations, which targeted UV-bright ($M_\text{UV}\lesssim-20$) candidate galaxies, yield a relatively high Ly$\alpha$ detection rate. This is in contrast to non/rare detection of Ly$\alpha$ reported in recent spectroscopic searches on UV-faint galaxies. This notable difference in a Ly$\alpha$ detection rate between UV-bright and -faint galaxies suggests an inhomogeneous nature of reionization in which reionization proceeds faster in overdense regions where bright galaxies are preferentially populated.
\end{enumerate}

Ly$\alpha$ is a major observational probe to trace the evolution of reionization. The presence of the clustered Ly$\alpha$ emitters reported in this work indicates that we are witnessing ionized regions in the IGM whereas the nondetections of Ly$\alpha$ from other sources reflect the neutral IGM. This is consistent with the general picture of reionization on the inhomogeneity in reionization.  Particularly, the LOS distribution of the $z\sim7.7$ cluster LAEs, in which the brightest galaxy is found in front of the others, suggests that the detailed analysis of Ly$\alpha$ observations on the IGM transmission can hint at outlining the scope of ionized regions around reionization-era galaxies as well as investigating how galaxies are distributed inside.

We caution that interpreting these results from Ly$\alpha$ observations remains challenging. This is mainly because current Ly$\alpha$ studies generally rely on the photometric selection of galaxies when their spectroscopic confirmations are not available, which inevitably includes some portion of low-redshift interloper galaxies. In addition, the lack of direct measurements of Ly$\alpha$ velocity offsets causes significant uncertainties in estimating the Ly$\alpha$ transmission from observations \citep[e.g.,][]{Mason2018a, Hoag2019a, Jung2020a, Jung2022a}.  Additionally, the size distribution of ionized bubbles also plays an important role in determining Ly$\alpha$ transmission at a fixed IGM neutral fraction \citep[e.g.,][]{Matthee2018a, Mason2020a, Park2021a, Qin2021a, Smith2021a}. These factors eventually compound the uncertainty of the final measurement of the neutral fraction of the IGM from Ly$\alpha$ observations. 

JWST observations, however, can place critical constraints on these uncertainties and improve the use of Ly$\alpha$ as a probe of reionization.  Specifically, even in the darkness of Ly$\alpha$, JWST can confirm the redshifts of numerous galaxies with non-Ly$\alpha$ emission lines or the Lyman-alpha break.  Also, Ly$\alpha$ velocity offsets can be measured directly from non-resonant emission lines.  Additionally, improved estimates of the ionizing photon production rate are possible via the use of nebular emission lines and/or better-constrained SED modeling \citep[e.g.,][]{Williams2022a, Robertson2022a}, which constrains the size of ionized bubbles around galaxies. In the new era of JWST, Ly$\alpha$ observations will eventually allow us to place strong constraints on the IGM neutral fraction during the epoch of reionization.

\acknowledgments
I.J. acknowledges support from NASA under award number 80GSFC21M0002. 
TAH is supported by an appointment to the NASA Postdoctoral Program (NPP) at NASA Goddard Space Flight Center, administered by Oak Ridge Associated Universities under contract with NASA.
This work was supported by a NASA Keck PI Data Award, administered by the NASA Exoplanet Science Institute. Data presented herein were obtained at the W. M. Keck Observatory from telescope time allocated to the National Aeronautics and Space Administration through the agency's scientific partnership with the California Institute of Technology and the University of California. The Observatory was made possible by the generous financial support of the W. M. Keck Foundation. The authors wish to recognize and acknowledge the very significant cultural role and reverence that the summit of Maunakea has always had within the indigenous Hawaiian community. We are most fortunate to have the opportunity to conduct observations from this mountain.

\appendix

\section{Spectroscopic Targets for Ly$\alpha$}
We list our spectroscopic targets in Table \ref{tab:targets} in order of decreasing photometric redshift, which includes the 3$\sigma$ rest-EW upper limits of Ly$\alpha$ for nondetection objects in the last column. 

\startlongtable
\begin{deluxetable*}{cccccccc}
\tablecaption{Summary of Spectroscopic Targets}
\tablehead{ \colhead{ID} & \colhead{R.A. (J2000.0)}         & \colhead{Decl. (J2000.0)}         & \colhead{$J_{\text{125}}$} & \colhead{$M_{\text{UV} }$\tablenotemark{a}} & \colhead{$z_{\text{phot}}$\tablenotemark{b}} & \colhead{$z_{\text{spec}}$\tablenotemark{c}} & \colhead{EW$_{\text{Ly}\alpha}$\tablenotemark{d} (\AA)} }
\startdata
{z8\_7364} & {215.035610} & { 52.892210} & { 25.6} & {-21.8}& {8.14$^{+0.56}_{-0.66}$} &{-} &{$<$35.7}\\
{z8\_62818} & {214.793960} & { 52.841540} & { 26.0} & {-21.1}& {8.04$^{+0.27}_{-0.29}$} &{-} &{$<$53.9}\\
{z8\_14498} & {214.943550} & { 52.845650} & { 26.6} & {-20.6}& {7.80$^{+0.72}_{-0.83}$} &{-} &{$<$92.5}\\
{z8\_70475} & {215.103630} & { 53.043030} & { 26.5} & {-20.7}& {7.79$^{+0.81}_{-1.48}$} &{-} &{$<$57.2}\\
{z8\_19326} & {215.119620} & { 52.982840} & { 27.0} & {-20.2}& {7.79$^{+0.75}_{-5.68}$} &{7.783} &{151.0$^{+125.4}_{-66.2}$}\\
{z8\_32350} & {214.999030} & { 52.941970} & { 25.3} & {-21.9}& {7.82$^{+0.72}_{-0.77}$} &{7.776} &{17.7$^{+8.6}_{-5.7}$}\\
{z7\_30645} & {215.095040} & { 53.014210} & { 25.2} & {-22.0}& {7.00$^{+0.36}_{-0.40}$} &{7.750} &{8.7$^{+4.3}_{-3.4}$}\\
{z7\_27591} & {215.132880} & { 53.047860} & { 26.3} & {-20.8}& {7.18$^{+0.58}_{-0.51}$} &{7.750} &{19.1$^{+9.0}_{-7.6}$}\\
{z8\_13573} & {215.150880} & { 52.989570} & { 26.5} & {-20.7}& {7.74$^{+0.72}_{-0.76}$} &{7.748} &{69.1$^{+29.8}_{-19.9}$}\\
{z7\_8626} & {215.114460} & { 52.951230} & { 26.3} & {-20.9}& {6.76$^{+0.36}_{-0.40}$} &{7.668} &{49.4$^{+17.5}_{-11.7}$}\\
{z8\_57340} & {215.100080} & { 53.072100} & { 26.2} & {-21.0}& {7.66$^{+0.72}_{-1.04}$} &{-} &{$<$36.6}\\
{z8\_35089} & {215.080330} & { 52.993230} & { 24.8} & {-22.4}& {7.65$^{+0.59}_{-0.60}$} &{-} &{$<$11.2}\\
{z7\_20237} & {215.106580} & { 52.975820} & { 26.1} & {-21.0}& {7.13$^{+0.81}_{-0.72}$} &{7.623} &{17.1$^{+8.6}_{-5.7}$}\\
{z8\_48797} & {215.136960} & { 53.001580} & { 26.6} & {-20.4}& {7.62$^{+0.57}_{-0.63}$} &{-} &{$<$86.8}\\
{z8\_47409} & {214.882250} & { 52.824670} & { 26.8} & {-20.4}& {7.61$^{+0.77}_{-1.35}$} &{-} &{$<$93.9}\\
{z8\_55956} & {214.737210} & { 52.818380} & { 26.6} & {-20.5}& {7.60$^{+0.73}_{-1.28}$} &{-} &{$<$78.3}\\
{z8\_52358} & {214.728630} & { 52.820880} & { 26.5} & {-20.6}& {7.59$^{+0.79}_{-5.31}$} &{-} &{$<$93.9}\\
{z8\_67892} & {214.880950} & { 52.891200} & { 26.4} & {-20.7}& {7.56$^{+0.64}_{-5.49}$} &{-} &{$<$93.7}\\
{z8\_21868} & {214.813040} & { 52.834230} & { 26.6} & {-20.4}& {7.53$^{+0.53}_{-1.01}$} &{-} &{$<$84.7}\\
{z7\_64424} & {215.131670} & { 53.076920} & { 26.1} & {-21.0}& {7.52$^{+0.51}_{-0.40}$} &{-} &{$<$30.7}\\
{z7\_13433} & {214.850830} & { 52.776660} & { 25.0} & {-22.1}& {7.11$^{+0.28}_{-0.26}$} &{7.478} &{22.2$^{+8.6}_{-7.0}$}\\
{z7\_31938} & {215.130040} & { 53.035510} & { 26.3} & {-20.8}& {7.46$^{+0.55}_{-0.54}$} &{-} &{$<$33.6}\\
{z7\_61615} & {214.995500} & { 52.987580} & { 26.5} & {-20.6}& {7.44$^{+0.74}_{-1.03}$} &{-} &{$<$67.7}\\
{z7\_63317} & {214.862990} & { 52.889430} & { 25.9} & {-21.1}& {7.42$^{+0.62}_{-0.67}$} &{-} &{$<$49.0}\\
{z7\_66460} & {214.990460} & { 52.971990} & { 26.0} & {-21.0}& {7.37$^{+0.63}_{-0.54}$} &{-} &{$<$33.0}\\
{z7\_17991} & {215.077870} & { 52.950110} & { 27.1} & {-19.9}& {7.36$^{+0.58}_{-0.67}$} &{-} &{$<$121.5}\\
{z7\_61983} & {215.132630} & { 53.084080} & { 25.9} & {-21.2}& {7.36$^{+0.37}_{-0.31}$} &{-} &{$<$22.8}\\
{z7\_12730} & {215.138580} & { 52.978710} & { 26.8} & {-20.3}& {7.35$^{+0.81}_{-5.72}$} &{-} &{$<$85.9}\\
{z7\_22848} & {215.115790} & { 53.045690} & { 25.9} & {-21.0}& {7.33$^{+0.61}_{-0.80}$} &{-} &{$<$25.2}\\
{z7\_68268} & {215.009710} & { 52.981390} & { 24.9} & {-22.2}& {7.30$^{+0.61}_{-0.78}$} &{-} &{$<$15.1}\\
{z7\_22554} & {215.132580} & { 53.058960} & { 27.4} & {-19.6}& {7.29$^{+0.89}_{-1.34}$} &{-} &{$<$108.3}\\
{z6\_39031\tablenotemark{e}} & {215.144960} & { 53.029710} & { 25.5} & {-21.4}& {7.64$^{+0.47}_{-0.45}$} &{-} &{-}\\
{z7\_16064} & {215.091040} & { 52.954280} & { 26.8} & {-20.2}& {7.25$^{+0.60}_{-0.80}$} &{-} &{$<$92.6}\\
{z7\_36800} & {214.797330} & { 52.788880} & { 26.8} & {-20.1}& {7.25$^{+0.80}_{-0.95}$} &{-} &{$<$118.5}\\
{z7\_33661} & {215.079120} & { 52.995750} & { 26.9} & {-19.9}& {7.25$^{+0.70}_{-0.60}$} &{-} &{$<$153.4}\\
{z7\_39792} & {214.941730} & { 52.884560} & { 26.3} & {-20.7}& {7.23$^{+0.51}_{-4.84}$} &{-} &{$<$111.5}\\
{z7\_27932} & {214.859170} & { 52.853590} & { 26.2} & {-20.7}& {7.14$^{+0.66}_{-5.74}$} &{-} &{$<$51.9}\\
{z7\_69794} & {215.077540} & { 53.026070} & { 26.0} & {-21.2}& {7.11$^{+0.32}_{-0.42}$} &{-} &{$<$25.7}\\
{z7\_12383} & {214.891540} & { 52.803070} & { 25.9} & {-21.1}& {7.11$^{+0.74}_{-5.33}$} &{-} &{$<$47.1}\\
{z7\_34392} & {214.946710} & { 52.900520} & { 26.6} & {-20.5}& {7.10$^{+0.39}_{-0.44}$} &{-} &{$<$105.4}\\
{z7\_60238} & {215.103540} & { 53.067080} & { 27.0} & {-20.0}& {6.97$^{+0.81}_{-1.28}$} &{-} &{$<$115.9}\\
{z7\_48468} & {215.068000} & { 52.953770} & { 25.9} & {-21.1}& {6.96$^{+0.29}_{-0.30}$} &{-} &{$<$37.2}\\
{z7\_64385} & {214.805040} & { 52.845870} & { 27.0} & {-19.6}& {6.92$^{+0.66}_{-5.23}$} &{-} &{$<$229.5}\\
{z7\_39204} & {214.828420} & { 52.810830} & { 25.0} & {-22.1}& {6.91$^{+0.28}_{-0.31}$} &{-} &{$<$14.5}\\
{z6\_40811} & {214.855170} & { 52.820750} & { 26.0} & {-21.0}& {6.76$^{+0.09}_{-5.68}$} &{-} &{$<$48.3}\\
{z6\_10540} & {214.979940} & { 52.861100} & { 25.5} & {-21.3}& {6.68$^{+0.54}_{-0.44}$} &{-} &{$<$43.8}\\
{z7\_18441} & {215.032080} & { 52.918970} & { 26.5} & {-20.2}& {6.66$^{+0.49}_{-0.73}$} &{-} &{$<$109.6}\\
{z7\_15372} & {214.987940} & { 52.879440} & { 25.1} & {-21.8}& {6.54$^{+0.12}_{-0.12}$} &{-} &{$<$22.5}\\
{z6\_20474} & {215.005970} & { 52.905310} & { 25.3} & {-21.6}& {6.49$^{+0.09}_{-5.04}$} &{-} &{$<$23.7}\\
{z6\_47325} & {215.026580} & { 52.927140} & { 26.1} & {-20.6}& {6.40$^{+0.30}_{-5.02}$} &{-} &{$<$43.5}\\
{z6\_12266} & {214.879170} & { 52.793910} & { 25.3} & {-21.5}& {6.39$^{+0.18}_{-0.33}$} &{-} &{$<$20.5}\\
{z6\_23620} & {215.162130} & { 53.077280} & { 25.6} & {-21.2}& {6.26$^{+0.13}_{-4.69}$} &{-} &{$<$21.1}\\
{z6\_24994} & {215.006790} & { 52.965040} & { 25.5} & {-21.3}& {6.22$^{+0.12}_{-0.13}$} &{-} &{$<$21.4}\\
{z6\_69545} & {214.984000} & { 52.960450} & { 25.2} & {-21.4}& {6.12$^{+0.30}_{-0.44}$} &{-} &{$<$25.1}\\
{z6\_12561} & {215.007900} & { 52.886100} & { 25.7} & {-21.0}& {6.11$^{+0.18}_{-4.85}$} &{-} &{$<$38.5}\\
{z6\_23791} & {215.049250} & { 52.997550} & { 25.0} & {-21.5}& {6.06$^{+0.11}_{-0.10}$} &{-} &{$<$14.8}\\
{z6\_30737} & {215.146750} & { 53.050340} & { 25.1} & {-21.6}& {6.05$^{+0.12}_{-0.12}$} &{-} &{$<$11.2}\\
{z6\_37712} & {214.790480} & { 52.781510} & { 25.0} & {-21.6}& {6.05$^{+0.15}_{-0.16}$} &{-} &{$<$20.2}\\
{z6\_5742} & {215.026260} & { 52.881630} & { 27.1} & {-}& {5.95$^{+0.69}_{-1.34}$} &{-} &{-}\\
{z6\_48598} & {214.987770} & { 52.896860} & { 27.0} & {-}& {5.84$^{+0.37}_{-0.50}$} &{-} &{-}\\
{z6\_66862} & {214.764580} & { 52.810830} & { 25.7} & {-}& {5.79$^{+0.25}_{-0.24}$} &{-} &{-}\\
\enddata
\tablenotetext{}{
$^{a}$$M_{\text{UV}}$ is estimated from the averaged flux over a 1450 -- 1550\AA\ bandpass from the best-fit galaxy SED model.\\
$^{b}$We present the 1$\sigma$ range of $z_{\text{phot}}$.\\
$^{c}$Spectroscopic redshifts are estimated from the detected Ly$\alpha$ emission lines.\\
$^{d}$$3\sigma$ upper limits, measured from the median flux limits from individual spectra. \\
$^{e}$ This object is not included in the analysis. We detected an emission line, but it is likely to be a low-redshift object from our SED fitting analysis.
}
\label{tab:targets}
\end{deluxetable*}

\bibliographystyle{aasjournal}
\bibliography{references}
\end{document}